\documentclass[useAMS,usenatbib]{mn2e} 
\usepackage{txfonts,graphicx} 
\usepackage{color} 
\usepackage[colorlinks=true,
            linkcolor=red,
            urlcolor=blue,
            citecolor=blue]{hyperref}
\usepackage{natbib,longtable,url} 
\def\arcsec{\hbox{$^{\prime\prime}$}}
\def\utw{\smash{\rlap{\lower5pt\hbox{$\sim$}}}}
\def\udtw{\smash{\rlap{\lower6pt\hbox{$\approx$}}}}

\def\diameter{{\ifmmode\mathchoice
{\ooalign{\hfil\hbox{$\displaystyle/$}\hfil\crcr
{\hbox{$\displaystyle\mathchar"20D$}}}}
{\ooalign{\hfil\hbox{$\textstyle/$}\hfil\crcr
{\hbox{$\textstyle\mathchar"20D$}}}}
{\ooalign{\hfil\hbox{$\scriptstyle/$}\hfil\crcr
{\hbox{$\scriptstyle\mathchar"20D$}}}}
{\ooalign{\hfil\hbox{$\scriptscriptstyle/$}\hfil\crcr
{\hbox{$\scriptscriptstyle\mathchar"20D$}}}}
\else{\ooalign{\hfil/\hfil\crcr\mathhexbox20D}}%
\fi}}

\def\bbbc{{\mathchoice {\setbox0=\hbox{$\displaystyle\rm C$}\hbox{\hbox
to0pt{\kern0.4\wd0\vrule height0.9\ht0\hss}\box0}}
{\setbox0=\hbox{$\textstyle\rm C$}\hbox{\hbox
to0pt{\kern0.4\wd0\vrule height0.9\ht0\hss}\box0}}
{\setbox0=\hbox{$\scriptstyle\rm C$}\hbox{\hbox
to0pt{\kern0.4\wd0\vrule height0.9\ht0\hss}\box0}}
{\setbox0=\hbox{$\scriptscriptstyle\rm C$}\hbox{\hbox
to0pt{\kern0.4\wd0\vrule height0.9\ht0\hss}\box0}}}}
\def\bbbq{{\mathchoice {\setbox0=\hbox{$\displaystyle\rm
Q$}\hbox{\raise
0.05\ht0\hbox to0pt{\kern0.4\wd0\vrule height0.9\ht0\hss}\box0}}
{\setbox0=\hbox{$\textstyle\rm Q$}\hbox{\raise
0.05\ht0\hbox to0pt{\kern0.4\wd0\vrule height0.9\ht0\hss}\box0}}
{\setbox0=\hbox{$\scriptstyle\rm Q$}\hbox{\raise
0.05\ht0\hbox to0pt{\kern0.4\wd0\vrule height0.8\ht0\hss}\box0}}
{\setbox0=\hbox{$\scriptscriptstyle\rm Q$}\hbox{\raise
0.05\ht0\hbox to0pt{\kern0.4\wd0\vrule height0.8\ht0\hss}\box0}}}}
\def\bbbt{{\mathchoice {\setbox0=\hbox{$\displaystyle\rm
T$}\hbox{\hbox to0pt{\kern0.25\wd0\vrule height0.95\ht0\hss}\box0}}
{\setbox0=\hbox{$\textstyle\rm T$}\hbox{\hbox
to0pt{\kern0.25\wd0\vrule height0.95\ht0\hss}\box0}}
{\setbox0=\hbox{$\scriptstyle\rm T$}\hbox{\hbox
to0pt{\kern0.25\wd0\vrule height0.95\ht0\hss}\box0}}
{\setbox0=\hbox{$\scriptscriptstyle\rm T$}\hbox{\hbox
to0pt{\kern0.25\wd0\vrule height0.95\ht0\hss}\box0}}}}
\def\bbbs{{\mathchoice
{\setbox0=\hbox{$\displaystyle\rm S$}\hbox{\raise0.5\ht0\hbox
to0pt{\kern0.38\wd0\vrule height0.45\ht0\hss}\hbox
to0pt{\kern0.52\wd0\vrule height0.5\ht0\hss}\box0}}
{\setbox0=\hbox{$\textstyle \rm S$}\hbox{\raise0.5\ht0\hbox
to0pt{\kern0.38\wd0\vrule height0.45\ht0\hss}\hbox
to0pt{\kern0.52\wd0\vrule height0.5\ht0\hss}\box0}}
{\setbox0=\hbox{$\scriptstyle \rm S$}\hbox{\raise0.5\ht0\hbox
to0pt{\kern0.38\wd0\vrule height0.45\ht0\hss}\raise0.05\ht0\hbox
to0pt{\kern0.52\wd0\vrule height0.45\ht0\hss}\box0}}
{\setbox0=\hbox{$\scriptscriptstyle\rm S$}\hbox{\raise0.5\ht0\hbox
to0pt{\kern0.38\wd0\vrule height0.45\ht0\hss}\raise0.05\ht0\hbox
to0pt{\kern0.52\wd0\vrule height0.45\ht0\hss}\box0}}}}
\def\bbbz{{\mathchoice {\hbox{$\sf\textstyle Z\kern-0.4em Z$}}
{\hbox{$\sf\textstyle Z\kern-0.4em Z$}}
{\hbox{$\sf\scriptstyle Z\kern-0.3em Z$}}
{\hbox{$\sf\scriptscriptstyle Z\kern-0.2em Z$}}}}

\def\aj{Astronomical Journal}
\def\araa{Annual Review of Astronomy and Astrophysics}
\def\apj{Astrophysical Journal}
\def\apjs{Astrophysical Journal, Supplement}
\def\aap{Astronomy and Astrophysics}
\def\aaps{Astronomy and Astrophysics, Supplement}

\def\mnras{Monthly Notices of the Royal Astronomical Society}
\def\pra{Physical Review A: General Physics}
\def\nat{Nature}%

\def\physscr{Physica Scripta}

\title[Non-LTE line formation of Fe in late-type stars IV]{Non-LTE line formation of Fe in late-type stars IV:  Modelling of the solar centre-to-limb variation in 3D} 
\author[Lind et al.]{K. Lind \thanks{E-mail: klind@mpia.de}$^{1,2}$, A.~M. Amarsi$^{1,3}$, M. Asplund$^3$, P.~S. Barklem$^2$, M. Bautista$^4$, M. Bergemann$^1$, 
\newauthor R. Collet$^5$, D. Kiselman$^6$, J. Leenaarts$^6$, T.~M.~D. Pereira$^7$\ \\
$^1$ Max-Planck Institute for Astronomy, K\"onigstuhl 17, 69117 Heidelberg, Germany \\ 
$^2$ Department of Physics and Astronomy, Uppsala University, Box 516, 751 20, Uppsala, Sweden \\ 
$^3$ Research School of Astronomy \& Astrophysics, Australian National University, Cotter Road, Canberra, ACT 2611, Australia \\
$^4$ Department of Physics, Western Michigan University, Kalamazoo, MI, 49008, USA\\
$^5$ Stellar Astrophysics Centre, Department of Physics and Astronomy, Aarhus University, Ny Munkegade 120, DK-8000 Aarhus C, Denmark\\
$^6$ Institute for Solar Physics, Dept. of Astronomy, Stockholm University, Albanova University Center, 106 91, Stockholm, Sweden\\
$^7$ Institute for Theoretical Astrophysics, University of Oslo, P.O.\,Box 1029 Blindern, N-0315 Oslo, Norway 
}

\begin{document}

\date{Accepted Date. Received Date 29.12.2016; in original Date}

\pagerange{\pageref{firstpage}--\pageref{lastpage}} \pubyear{2011}

\maketitle

\label{firstpage}

\begin{abstract} {Our ability to model the shapes and strengths of iron lines in the solar spectrum is a critical test of the accuracy of the solar iron abundance, which sets the absolute zero-point of all stellar metallicities. We use an extensive 463-level Fe atom with new photoionisation cross-sections for FeI as well as quantum mechanical calculations of collisional excitation and charge transfer with neutral hydrogen; the latter effectively remove a free parameter that has hampered all previous line formation studies of Fe in non-local thermodynamic equilibrium (NLTE). For the first time, we use realistic 3D NLTE calculations of Fe for a quantitative comparison to solar observations. We confront our theoretical line profiles with observations taken at different viewing angles across the solar disk with the Swedish 1-m Solar Telescope. We find that 3D modelling well reproduces the observed centre-to-limb behaviour of spectral lines overall, but highlight aspects that may require further work, especially cross-sections for inelastic collisions with electrons. Our inferred solar iron abundance is $\log(\epsilon_{\rm Fe})=7.48\pm0.04\rm\,dex$.} \end{abstract}


%
%
\section{Introduction} 
Because of the dominance of lines of atomic iron in the spectra of cool stars, the iron abundance is often used as a proxy for total metal content, or metallicity. Neutral and singly ionised iron with different properties are also frequently used for determination of spectroscopic stellar parameters. This makes [Fe/H] arguably the most important abundance indicator when studying the evolution of stars and galaxies. The iron abundance of the Sun itself is important not only as an anchor for the cosmic [Fe/H]-scale, but it also influences the structure and evolution of stars because it is a large opacity contributor \citep[][and references therein]{Bailey15}. 

It has long been known that the ionisation balance of FeI--FeII departs from local thermodynamic equilibrium (LTE) in the photospheres of late-type stars \citep[e.g.][]{Athay72,Rutten84,Thevenin99, Korn03}. However, non-LTE (hereafter NLTE) calculations of neutral iron have suffered from large systematic uncertainties due to poorly constrained atomic data, in particular the efficiency of collisions with electrons and neutral hydrogen \citep[e.g.][]{Mashonkina11b}. Further, the large complexity of the atomic structure of iron has prevented consistent NLTE calculations to be performed with realistic 3D radiation-hydrodynamical simulations of solar and stellar photospheres. 

In Paper I and II of this publication series \citep{Bergemann12,Lind12a}, we presented a model atom for iron with the efficiency of H collisions calibrated using high-quality spectroscopic observations of well-studied benchmark stars, including the Sun. We employed both standard 1D atmospheric models and so called average-3D models (hereafter $\rm\langle3D\rangle$), which are spatial and temporal averages of full 3D radiation-hydrodynamical simulations of stellar atmospheres. In Paper III \citep{Amarsi16b} we presented a new model atom including quantum mechanical calculations of hydrogen collisions (Barklem, in prep.) and demonstrated its performance for metal-poor benchmark stars using full 3D NLTE calculations. Here we further improve the atom and confront our 3D NLTE predictions with the observed centre-to-limb variation of iron lines in Sun. 

\citet{Nordlund84,Nordlund85} pioneered the investigation of NLTE line formation of iron in 3D hydrodynamical model atmospheres more than three decades ago. The first paper studied the departure of FeI--FeII from Saha ionisation balance and reported significant $\rm(0.2\,dex)$ over-ionisation of the neutral species. The second paper used a two-level FeI atom, coupled to a FeII continuum, and predicted significant line weakening of the example FeI line at 5225\AA\  due to a superthermal source function. 

\citet{Shchukina01} later studied NLTE line formation in a hydrodynamical model of the Sun in the so called 1.5D approximation, neglecting horizontal radiative transfer. They used a 248 level FeI+FeII atom and concluded that NLTE effects vary strongly with the granulation pattern and the FeI line properties, with a net NLTE correction to FeI line abundances of up to $+0.12$\,dex for the lowest-excitation lines. The only previous work investigating NLTE line formation of iron in the Sun using a multi-level atom and full 3D radiative transfer is the series by \citet{Holzreuter12,Holzreuter13,Holzreuter15}, in which the authors rigorously compare synthetic line profiles generated under different assumptions. However, they were limited to using a strongly simplified 23-level atom and made no quantitative comparison to observations. These earlier studies have in common that they included only experimentally known energy levels of iron and neglected the influence of hydrogen collisions on the statistical equilibrium, both of which exaggerate the NLTE effects.  

We present full 3D NLTE calculations using a comprehensive 463 level atom with realistic atomic data to enable a direct comparison to the most constraining observations possible, i.e. high-spectral resolution and high signal-to-noise (S/N) observations of the Sun at different viewing angles. The paper is divided in the following sections: Sect.\,\ref{sect:method} outlines the observations, the assembly and reduction of the model atom, and the method used for spectral synthesis.  Sect.\,\ref{sect:results} presents the results for the solar centre-to-limb variation of iron lines and the solar iron abundance. Sect.\,\ref{sect:conc} summarises our conclusions.  

\section{Method}
\label{sect:method}

\begin{figure} 
\begin{center} 
\includegraphics[scale=0.43,viewport=3cm 1cm 21cm 21cm]{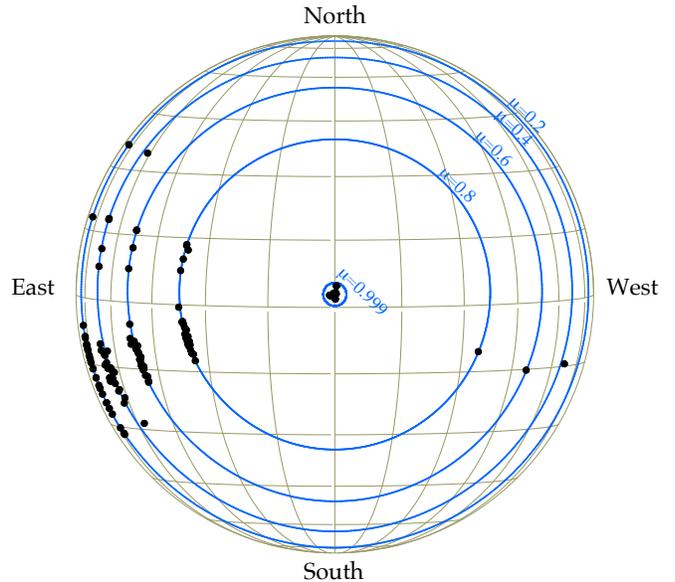}
\caption[]{Overview of the SST pointings on the solar disk, inclined by the heliographic latitude of the observer.  The blue circles mark the targeted $\mu$-angles and $\mu=0.999$ for reference.} 
\label{fig:pos} 
\end{center} 
\end{figure}

\subsection{Observations}  
We acquired spectroscopic data with high spatial and spectral resolution using the TRIPPEL \citep{Kiselman11} instrument at the Swedish 1-m Solar Telescope \citep[SST,][]{Scharmer03} on La Palma. The observing campaign lasted from 23 June to 8 July, 2011. 

Three spectrographic cameras and three imaging cameras were operated simultaneously.  Three  different setups were used, resulting in a total of nine spectral windows with wavelength bands specified in Table \ref{tab:bands}. Two slit-jaw cameras recorded simultaneous images at approximately 5320\AA\, and 6940\AA\,, respectively. The third camera was used to monitor the magnetic activity of the region with a 1.1\AA\ filter centered on the Ca\,II H line. Five different heliocentric angles on the solar disk were targeted, corresponding to $\mu\equiv\cos{\theta}=0.2, 0.4, 0.6, 0.8$ and $1.0$, where $\theta$ is the angle between the ray direction and the surface normal. The number of observations at each pointing is listed in Table \ref{tab:bands}, discarding exposures that failed due to suspected tracking problems (usually due to very bad seeing), or regions with obvious activity as deemed from the Ca\,II~H core emission.

The intensity contrast peaks at disk centre and exposures were made while scanning the spectroscopic slit over a small region in order to reduce the imprint of the local granulation pattern. At other pointings, the slit position was held fixed and aligned parallel to the closest part of the solar limb. The telescope field rotation caused the actual position selected in this way to depend on the time of day, as is evident in Fig.~\ref{fig:pos}. Of the two possible choices for a specific $\mu$ value and time of day, the one showing the least activity was preferred. 

For the $\mu=0.2$ pointings, the position of the slit could be measured accurately using the slit-jaw images, which include the solar limb. For the other pointings, $\mu$ was determined from the output of the telescope tracking system. In order to get the readings as accurate as possible, frequent calibrations by pointing at the limb at four position angles to find the solar centre were made. Somewhat conservatively, we estimate the accuracy in the the nominal values to be $\pm 15\arcsec$.

As evident from Fig. \ref{fig:pos} and Table \ref{tab:bands}, the pointing accuracy as expressed in $\mu$ degrades with decreasing $\mu$ value, i.e. from the centre towards the limb. The $\mu=0.2$ pointing deviates from this trend since it was determined from the slit-jaw images and the largest uncertainties thus affect the $\mu=0.4$ observations. In total we kept 147 pointings, 4-20 for each $\mu$-value and configuration, as detailed in Table \ref{tab:bands}. For each $\mu$ pointing and wavelength, Table~\ref{tab:bands} gives the mean value of $\mu$ and its error estimate calculated from the standard deviation of the nominal position readings combined with a systematic error. For $\mu\ge0.4$ the systematic component was computed using the $15\arcsec$ calibration error and for $\mu=0.2$, we used the approximate spatial extent of the slit in the same way as \citet{Pereira09a}. 

The data were reduced using the same method and software as described in \citet{Pereira09a}. First, the data were corrected for dark current and flat fielded using calibration exposures taken in close connection to the observations. 
Geometrical distortions in the spectrograms were then removed using polynomial fits to the location of selected spectral lines (for smile) and with the help of a grid place across the slit (for keystone). Wavelength calibration was made by cross-correlation of spectral lines with the disk-centre FTS atlas of Brault \& Noyes (1987)\footnote{\url{ftp://ftp.hs.uni- hamburg.de/pub/outgoing/FTS- Atlas}}. The same atlas was used to model the internal straylight of the spectrograph (assumed constant over each spectrogram) as well as systematic spectral artefacts that the flat-fielding cannot correct for. The result of the reduction procedure is to force mean spectra from the quiet disk centre to be as close as possible to the reference atlas spectrum. The same corrections are then applied to all spectra.

In this paper, we analyse the centre-to-limb behaviour of iron lines and create a single 1D spectrum for each pointing by coadding the individual spectrograms and then forming an average along the slit direction. This increases the S/N and, following \citet{Pereira09b}, removes the need for Fourier filtering of photon noise that was applied by \citet{Pereira09a}. The S/N per pixel of the average spectra ranges between 1000--4000 at a spectral resolution of $\lambda/\delta\lambda\approx150,000$. The S/N ratio was estimated from the median standard deviation of all measurements at a given wavelength.

\begin{table*}
      \caption{Summary of the observational configuration. Columns A-C give the wavelength band of each of the three spectrographic cameras. $\#$ represents the number of pointings.}
         \label{tab:bands}
         \centering
         \begin{tabular}{llllllllllllll}
                \hline\hline
                 Set & A & B & C &  $\#$ & $\mu$ &  $\#$ & $\mu$  &  $\#$ & $\mu$  &  $\#$ & $\mu$  &  $\#$ & $\mu$ \\
                 & [\AA\ ]& [\AA\ ]& [\AA\ ]   \\
                                \hline
	      1	& 5366-5377 &  6147-6159 & 8710-8728 & 6 	& 0.201 		& 4 		& 0.380	      	& 7 	& 0.600     	& 7 	&  0.802 		& 4 		& 1.0000 			\\
	         &     		      &                       &                      &	& $\pm0.007$ 	&   		& $\pm0.032$	&  	& $\pm0.014$	&  	& $\pm0.005$ 	&  		& $\pm0.0005$		\\
	      2 & 5378-5390 &  6159-6172 & 8727-8744 & 7 	& 0.205		& 7 		& 0.393	  	& 7 	& 0.604	   	& 6 	&  0.803	 	& 12 		& 1.0000	 		\\
	         &     		      &                       &                      &	& $\pm0.005$ 	&   		& $\pm0.019$	&  	& $\pm0.009$	&  	& $\pm0.005$ 	&  		& $\pm0.0003$		\\
	      3 & 8656-8668 &  7825-7842 & 8691-8708 & 16	& 0.203	 	& 16		& 0.397		& 19 & 0.603	  	& 20 	&  0.801	 	& 9 		& 1.0000	 		\\ 
	      	 &     		      &                       &                      &	& $\pm0.006$ 	&   		& $\pm0.027$	&  	& $\pm0.006$	&  	& $\pm0.004$ 	&  		& $\pm0.0005$		\\	      	      
                   \hline 
         \end{tabular}
\end{table*}

\begin{figure} 
\begin{center} 
\includegraphics[scale=0.33,viewport=2cm 1cm 25cm 21cm]{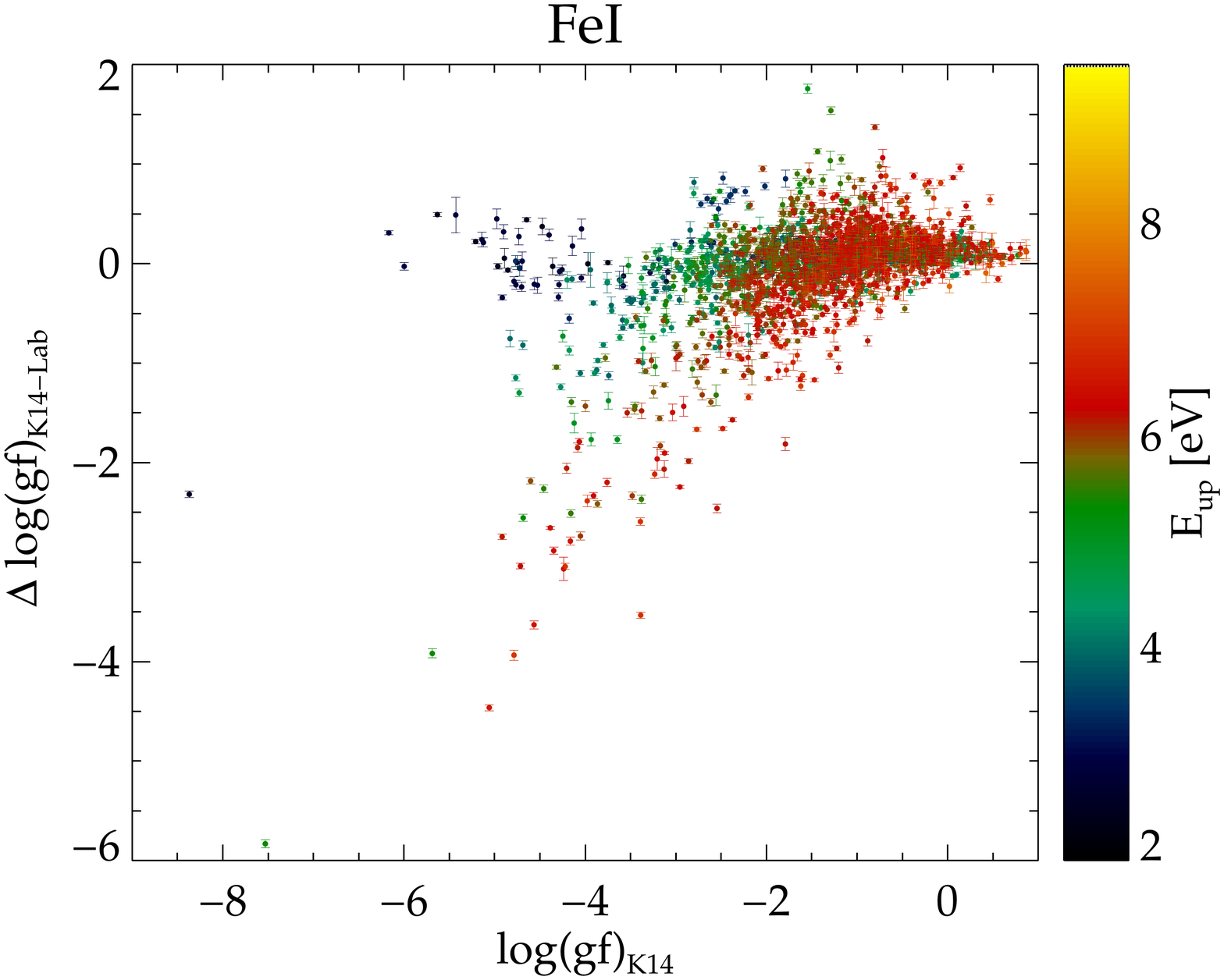}
\includegraphics[scale=0.33,viewport=2cm 0cm 25cm 21cm]{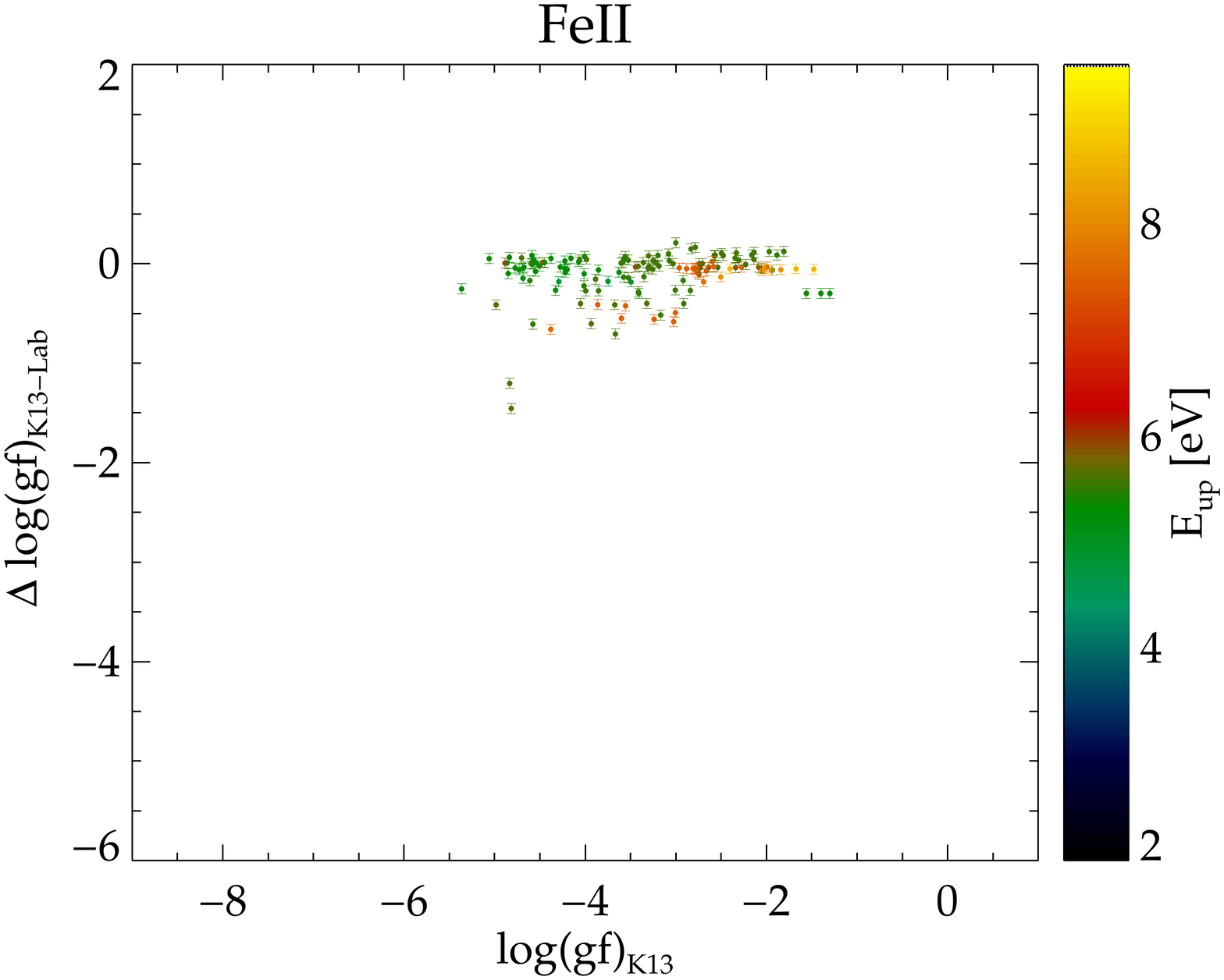}
\caption[]{Comparison between theoretically predicted \citep{K13,K14} and experimentally measured (see text for references) oscillator strengths of FeI and FeII. The experimental uncertainties are plotted as error bars. For FeII, a representative experimental uncertainly was set to $0.05$\,dex.} 
\label{fig:fvalues} 
\end{center} 
\end{figure}

\subsection{Atomic data} 
\label{sect:atomdata}

The non-LTE calculations are performed by iterative solutions of the radiative transfer and statistical equilibrium equations, until the level populations have converged at all points in the atmosphere. The statistical equilibrium solution requires knowledge of the relevant radiative and collisional transition probabilities, which are collected in a model atom. The literature sources and databases used to assemble the model atom were listed in \cite{Amarsi16b}. In this section we reiterate the main points and present more detail. 

\begin{table*}
      \caption{Atomic data for the iron lines used for the centre-to-limb analysis. The $W_\lambda$ columns list the equivalent widths measured for the five different $\mu$-angles using direct integration over the wavelength range specified by $\lambda_{\rm int}$.}
         \label{tab:lines}
         \centering
         \begin{tabular}{lllllllllllllc}
                \hline\hline
		Ion & $\lambda_{\rm air}$ & $E_{\rm low}$ & $\log(gf)$ & $\log(\gamma)$ & $\sigma^{(b)}$& $\alpha^{(b)}$& $C_4^{(c)}$ & \multicolumn{5}{c}{$W_{\lambda}$ [m\AA ]}   &$\lambda_{\rm int.}$\\
                          & [\AA\ ]          & [eV]                   &                   &   Rad.\,$^{(a)}$   &                   &                          &                        &    $\mu=1.0$  &    $\mu=0.8$  &    $\mu=0.6$  &    $\mu=0.4$  &    $\mu=0.2$  &  [\AA ] \\
                                \hline
FeI & 5367.4659  	& 4.415	&   0.443$^{(d)}$   	& 8.32 	&    972 & 0.280	&  -13.11 &   168.5  &   166.3   &  164.4  &   158.7   &  143.4   & $5367.10-5368.10$ \\ 
&      &            		&     		&                			& 		&    	&         &  $\pm1.3$  & $\pm1.3$  & $\pm1.4$  & $\pm1.2$  & $\pm1.5$ \\
FeI & 5373.7086  & 4.473&  -0.710$^{(e)}$   & 8.13 &   1044 & 0.282  &  -13.76 &    59.4  &    58.4   &   58.5  &    58.5   &   54.6   & $5373.62-5373.82$ \\ 
 &   &           &     &                & &            &         &  $\pm0.6$  & $\pm0.4$  & $\pm0.6$  & $\pm0.6$  & $\pm0.7$ \\
FeI & 5379.5736  & 3.695&  -1.514$^{(d)}$   & 7.85 &    363 & 0.249  &  -15.51 &    62.8  &    63.1   &   64.2  &    66.4   &   66.1   & $5379.20-5379.72$\\ 
 &   &           &     &                & &            &         &  $\pm0.6$  & $\pm0.7$  & $\pm0.6$  & $\pm0.9$  & $\pm0.8$ \\
FeI & 5383.3685  & 4.313&   0.645$^{(d)}$   & 8.30 &    836& 0.278  &  -13.83 &   219.3  &   217.7   &  214.9  &   208.3   &  188.4   & $5382.70-5384.00$\\ 
 &   &           &     &                & &            &         &  $\pm1.8$  & $\pm1.9$  & $\pm1.6$  & $\pm1.4$  & $\pm1.6$ \\
FeI & 5386.3331  & 4.154&  -1.670$^{(f)}$   & 8.45 &    930& 0.278  &  -13.02 &    31.0  &    31.9   &   33.1  &    34.7   &   34.8   & $5386.10-5386.45$\\ 
 &   &           &     &                & &            &         &  $\pm0.3$  & $\pm0.4$  & $\pm0.3$  & $\pm0.5$  & $\pm0.4$ \\
FeI & 5389.4788  & 4.415&  -0.418$^{(g)}$   & 8.32 &    959& 0.280  &  -13.53 &    87.9  &    86.8   &   86.2  &    85.4   &   80.4   & $5389.30-5389.65$\\ 
 &   &           &     &                & &            &         &  $\pm0.6$  & $\pm0.6$  & $\pm0.7$  & $\pm0.9$  & $\pm0.7$ \\
FeII& 6149.2459  & 3.889&  -2.840$^{(h)}$   & 8.50 &    186& 0.269  &  -16.11 &    38.3  &    38.4   &   37.2  &    36.7   &   31.4   & $6149.05-6149.40$\\ 
 &   &           &     &                & &            &         &  $\pm0.3$  & $\pm0.5$  & $\pm0.6$  & $\pm0.6$  & $\pm0.5$ \\
FeI & 6151.6173  & 2.176&  -3.299$^{(i)}$   & 8.29 &    277& 0.263  &  -15.55 &    49.0  &    49.6   &   52.2  &    55.4   &   55.8   &$6151.30-6151.85$ \\ 
 &   &           &     &                & &            &         &  $\pm0.8$  & $\pm0.7$  & $\pm0.5$  & $\pm0.7$  & $\pm1.6$ \\
FeI & 6157.7279  & 4.076&  -1.160$^{(f)}$   & 7.89 &    375& 0.255  &  -15.36 &    62.6  &    61.6   &   62.4  &    63.2   &   60.2   & $6157.50-6157.85$\\ 
 &   &           &     &                & &            &         &  $\pm0.7$  & $\pm0.6$  & $\pm0.7$  & $\pm0.9$  & $\pm1.2$ \\
FeI & 6165.3598  & 4.143&  -1.473$^{(d)}$   & 8.00 &    380& 0.250  &  -15.34 &    44.5  &    44.8   &   45.2  &    45.9   &   44.6   & $6165.25-6165.55$\\ 
 &   &           &     &                & &            &         &  $\pm0.4$  & $\pm0.5$  & $\pm0.4$  & $\pm0.6$  & $\pm0.6$ \\
FeI & 8699.4540  & 4.956&  -0.370$^{(e)}$   & 8.74 &    817& 0.272  &  -14.59 &    73.7  &    72.4   &   70.7  &    68.0   &   61.9   & $8699.20-8699.85$\\ 
 &   &           &     &                & &            &         &  $\pm0.8$  & $\pm0.9$  & $\pm1.0$  & $\pm1.0$  & $\pm0.9$ \\
              \hline
              \multicolumn{14}{l}{$^{(a)}$ Radiative broadening is given by the logarithm (base 10) of the FWHM given in $\rm rad\,s^{-1}$.} \\
              \multicolumn{14}{l}{$^{(b)}$ \citet{Anstee95} notation for the broadening cross-section ($\sigma$) for collisions by H\,I at 10\,$\rm km\,s^{-1}$ and its velocity dependence ($\alpha$).} \\
             \multicolumn{14}{l}{$^{(c)}$ Stark broadening constant.} \\
               \multicolumn{14}{l}{$^{(d)}$ \citet{BWL}, $^{(e)}$ \citet{2014MNRAS.441.3127R}, $^{(f)}$ \citet{MRW}, $^{(g)}$ \citet{FMW}, $^{(h)}$ \citet{RU}, $^{(i)}$ \citet{GESB82c}.} \\
         \end{tabular}
\end{table*} 

\subsubsection{Energy levels}

Energy levels were downloaded from Robert Kurucz's online database, updated in 2013 for FeII \footnote{\url{http://kurucz.harvard.edu/atoms/2601}} and 2014 for FeI \footnote{\url{http://kurucz.harvard.edu/atoms/2600}}. These data are referenced in the VALD3 \citep{Ryabchikova15} data base as \citet[]["K13"]{K13} and \citet[]["K14"]{K14}, respectively, and include both observed and theoretically predicted energy levels. The importance of the inclusion of predicted energy levels was demonstrated by \citet{Mashonkina11b}. There are 2,980 energy levels of FeI below the first ionisation potential ($63,737\,\rm cm^{-1}=7.902\,eV$), approximately two thirds of which have not been observed. For FeII we consider the 116 energy levels below 60,000\,$\rm cm^{-1}$, all of them observed, as more highly excited levels  are not relevant in late-type stellar atmospheres (the second ionisation potential of Fe is $130,655\,\rm cm^{-1}=16.199\,eV$). 

We have homogenised the nomenclature of electron configurations and terms to enable energy levels to be merged. Energies with terms given in jj-coupling notation, e.g. "2+[1+]" have instead been designated by the leading eigenvector's term in LS-coupling notation, e.g. "7F".  This convention is used in the creation of the term diagrams shown in Fig.\,\ref{fig:Fe804} and Fig.\,\ref{fig:Fe463}.

\subsubsection{Transition probabilities}

The \citet{K13,K14} database contains 533,772 radiative transitions between bound levels of FeI and 1174 between the bound levels we consider for FeII. We have cross-referenced these data with laboratory measurements of transition probabilities carried out since the late 1970's and identified 2080 matches  (0.4\% of all lines) for FeI and 115 matches for FeII (10\% of all lines). The references used for FeI are \citet{BIPS,GESB79b,GESB82c,GESB82d,GESB86,BKK,BK,BWL,2014ApJS..215...23D,2014MNRAS.441.3127R} and for FeII we use the re-normalised compilation by \citet{Melendez10}. The source with smallest quoted uncertainty was adopted for lines with multiple sources. 

Fig.\,\ref{fig:fvalues} compares theoretical and experimental values of $\log(gf)$ for both ionisation states. We find that the agreement is typically better for strong transitions; for $\log(gf)_{\rm K14}>-2$, theoretical values for FeI show a bias and scatter with respect to experiment of $0.08\pm0.29$\,dex, which increases in magnitude and changes sign to $-0.40\pm0.79$\,dex at $\log(gf)_{\rm K14}<-2$. For weak lines, there appears to be a correlation with the energy of the upper level involved in the transitions, such that the disagreement is very strong for lines with highly excited upper energy levels, while the least excited are in as good agreement with theory as stronger lines. For FeII lines, we find a bias of $-0.11\pm0.24$\,dex. The comparison suggests that the use of theoretical data for diagnostic lines should be avoided for precision spectroscopy. However, sensitivity tests that we carried out indicate that Fe NLTE level populations in the Sun are not sensitive to uncertainties in oscillator strengths of this magnitude. All lines selected for abundance analysis in Sect. \ref{sect:feabund} have laboratory measurements of $\log(gf)$.

\subsubsection{Photo-ionisation cross-sections}

We computed total and partial (state-to-state) photoionisation cross-sections for Fe I with the R-matrix method for atomic scattering as implemented in the RMATRX package \citep{Berrington95}.  These calculations employed close coupling expansion of 157 states of the Fe II target ion from 35 configurations made by atomic orbitals up to principal quantum number $n=6$. The atomic dataset includes cross-sections for 936 LS terms of Fe I with $n\le 10$ and $l\le 7$. Details of this calculation will be presented elsewhere (Bautista \& Lind 2016, in preparation). This calculation is considerably larger and more accurate than our previous computations of atomic data in \citet{Bautista97}. We use the total, not partial, photoionisation cross-sections in our model atom to limit the number of bound-free transitions.  Each FeI level is thus bound to a single FeII level, as shown in Fig.\,\ref{fig:Fe463}.


\begin{figure*} 
\begin{center} 
\includegraphics[scale=0.67,viewport=2cm 10.5cm 26cm 20.0cm]{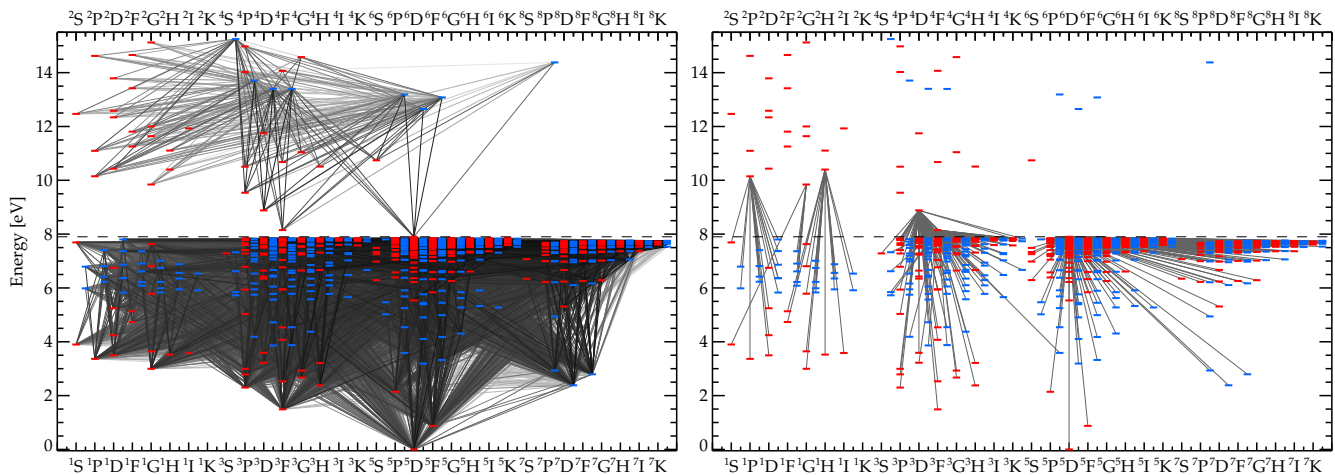}
\caption[]{The complete Fe model atom without fine structure. FeI levels are shown below the dashed line, which indicates the first ionisation potential, and the associated terms are listed at the bottom x-axis. The FeII levels considered in this work are shown above the dashed line and the associated terms are listed at the top axis. Even parity terms are displayed in red and odd parity terms in blue. The left-hand panel shows all radiative bound-bound transitions and the right-hand panel shows all bound-free transitions.} 
\label{fig:Fe804} 
\end{center} 
\end{figure*}

\begin{figure*} 
\begin{center} 
\includegraphics[scale=0.67,viewport=2cm 10.5cm 26cm 20.0cm]{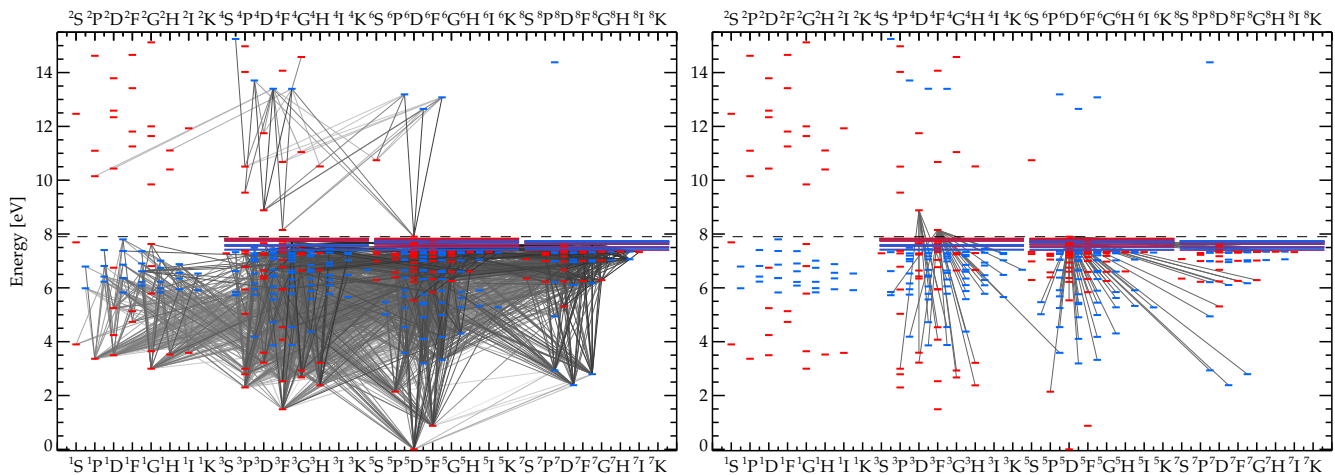}
\caption[]{Same as in Fig.\,\ref{fig:Fe804}, but for the reduced model atom used for 3D NLTE calculations. Merged levels are indicated with longer horisontal lines.} 
\label{fig:Fe463} 
\end{center} 
\end{figure*}

\begin{figure*} 
\begin{center} 
\includegraphics[scale=0.60]{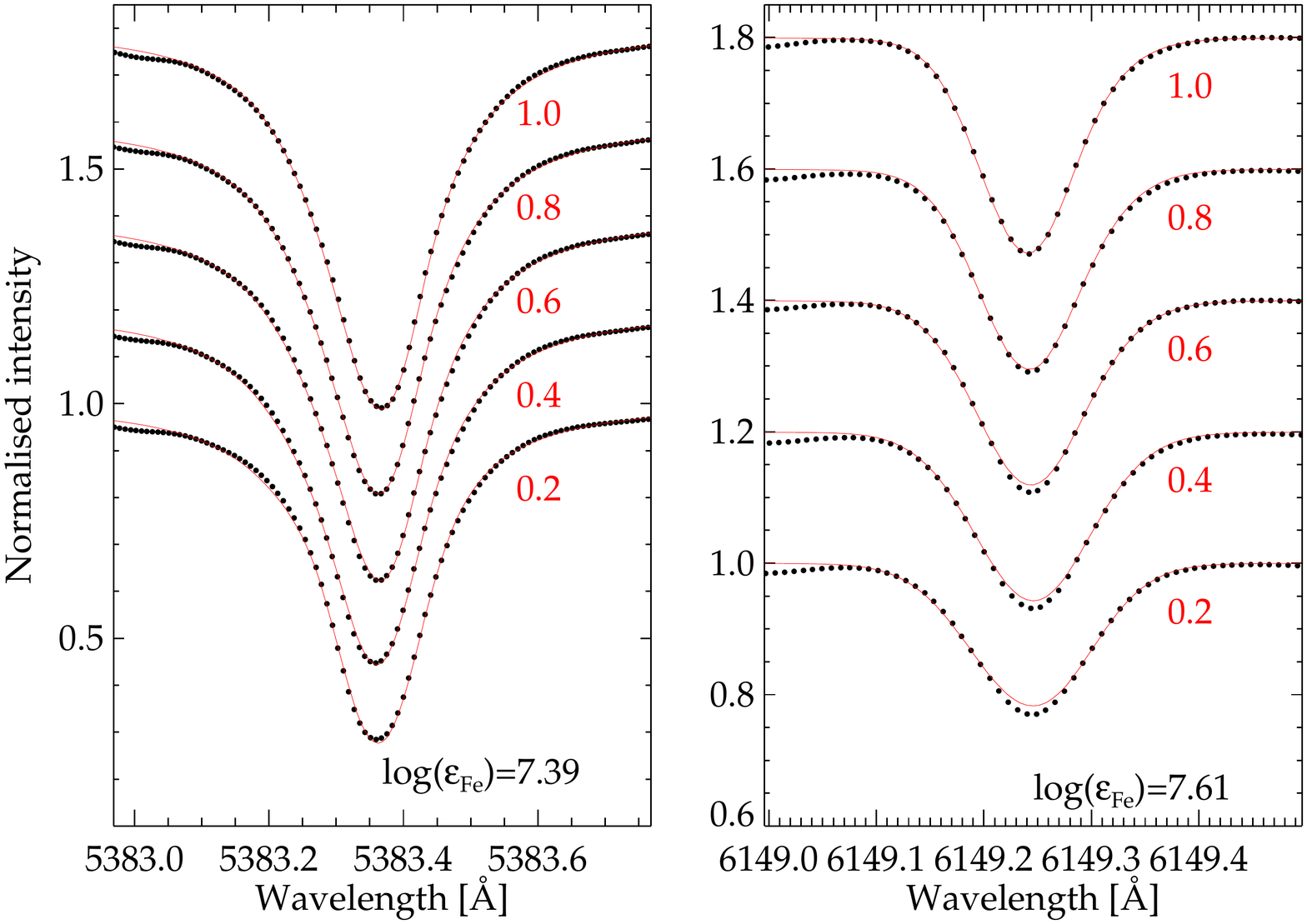}
\caption[]{Normalised observed (bullets) and synthetic (red lines) centre-to limb profiles for two iron lines, where the numbers below each spectrum correspond to the approximate $\mu$-angle. The synthetic line profiles have been computed in 3D NLTE and the iron abundance has been calibrated for each line to match the disk centre intensity. The calibrated abundance used for synthesis is indicated at the bottom of each panel. Both observed and synthetic spectra have been radial-velocity corrected so that the line centres coincide with the rest wavelength. Spectra for $\mu\ge0.4$ have been incrementally offset vertically by $+0.2$. } 
\label{fig:clvprof} 
\end{center} 
\end{figure*}

\subsubsection{Electron collisions}
We adopt the results of \citet{Zhang95}, who used the R-matrix method to compute collision rates between electrons and 18 low-excitation states of singly ionised Fe. When not available for bound-bound and bound-free electron impact collisions, we follow the semi-empirical recipes given by \citet{Allen00} for FeI and FeII, which are originally from \citet{vanRegemorter62} and \citet{Bely70}. The same formula was used for optically allowed and forbidden transitions, assuming $f=0.005$ for the latter, which gives the two types of transitions similar efficiencies. A comparison between rate coefficients computed by van Regmorter and \citet{Zhang95} for bound-bound FeII transitions gives a root mean square deviation of $0.6$\,dex in the temperature interval $3,000-10,000$\,K. For bound-free transitions, \citet{Allen00} mentions a probable uncertainty of 0.3\,dex. We note that more recent collisional data for FeII now exist and should be used for NLTE calculations \citep{Bautista15}. For the Sun, NLTE effects on FeII lines are insignificant, so the new data would not influence our results. 
 
\subsubsection{Hydrogen collisions} 
Collision rates for excitation processes, Fe($\alpha ^{2S+1}L$) + H($1s$) $\rightarrow$ Fe($\alpha'^{2S'+1}L'$) + H($1s$), and charge transfer processes, Fe($\alpha^{2S+1}L$) + H($1s$) $\rightarrow$ Fe$^+$($\alpha'^{2S'+1}L'$) + H$^-$, due to low-energy hydrogen atom collisions on neutral iron have been calculated with the asymptotic two-electron method presented by \citet{Barklem16}. The calculation used here includes 138 states of FeI, and 11 cores of FeII, leading to the consideration of 17 symmetries of the FeH molecule.  These data will be the subject of a future publication (Barklem, in prep.).

For transitions with no data available, we approximated values using robust fits to the behaviour of the (logarithmic) quantum mechanical rate coefficients with transition energy at a given temperature. Linear fits were used for de-excitation rates and second order polynomials were used for charge exchange rates. The dispersion around the fits are approximately 1.2\,dex in the temperature interval $3,000-10,000$\,K. A more elaborate discussion about the appropriate functional forms of such fits is given by Ezzeddine et al. (submitted).

\subsection{Atom reduction}
\label{sect:atomred}

In its complete form, our Fe model atom contains more than 3,000 fine-structure energy levels, coupled by half a million radiative transitions. To establish the statistical equilibrium using this atom would mean having to solve the radiative transfer equation for at least hundreds of thousands of frequency points, which is not feasible in 3D. The atom must therefore be simplified, while preserving the overall NLTE behaviour.

The traditional method used to reduce the size of complex model atoms is to merge close energy levels, implicitly assuming that the levels have the same departure coefficients. The degeneracies of the levels that are merged are used as weights in the calculation of the mean energy and the radiative transition probabilities corresponding to the merged level, in such a way that the sum of $gf$ is preserved. We start by following this approach for the collapse of the fine-structure levels, resulting in 762 bound levels of FeI, 41 levels of FeII, and the FeIII ground state. These levels are coupled by 92,567 transitions between bound states of FeI and 226 transitions between bound states of FeII. All FeI levels are coupled to a core FeII state and the photoionisation cross-sections are tabulated over 1,000-2,000 frequency points each. This model will be used as reference model atom and its term diagram is illustrated in Fig.\,\ref{fig:Fe804}. The simplification process has so far preserved level configuration, term, and parity for all levels.

We thereafter proceed to test how much further the atomic level structure can be simplified without causing a significant change in the departure coefficients. We use the $\rm\langle3D\rangle$ structure of the Sun as the default test model in this section, adopting a depth-independent microturbulence value of $1\rm\,km\,s^{-1}$. Above a certain energy limit, FeI energy levels are now merged that share the same multiplicity, parity, and configuration. We gradually decrease this energy limit, while monitoring the difference in equivalent width with respect to the reference model atom, for lines between $200$\,nm and $2\rm\mu m$. The number of levels were thereby reduced from 804 to 463. 

If all radiative transitions were kept, the 463 level atom would still contain approximately 37,000 transitions. However, many transitions do not contribute significantly to make the level populations depart from LTE. To reduce the number of transitions and enable full 3D calculations, we first computed the net radiative imbalance for each transitions in the $\rm\langle3D\rangle$ model of the Sun, assuming LTE populations, i.e., $\Delta_{ij}=|n_{i}R_{ij}-n_jR_{ji}|$. We then selected a point in the atmosphere ($\tau_{\rm500\,nm}\approx0.01$), where the NLTE effects are noticeable and relevant for line formation, and removed radiative transitions with a relatively small value of $\Delta_{ij}$. Thereby, only 3,000 bound-bound transitions and 100 bound-free transitions were kept. We note that the choice of reference depth does not strongly influence which transitions are discarded. Finally, the wavelength grids of the photo-ionisation cross-sections were down-sampled heavily, to a factor 30 fewer points. The final reduced atom contains approximately 17,000 frequency points and preserves $\rm\langle3D\rangle$ equivalent widths for the Sun within 0.01\,dex, compared to the reference atom. We note that, within these small uncertainties, the smaller atom gives slightly less efficient over-ionisation of FeI than the larger atom, but that further merging of energy levels would have the opposite effect because the collisional coupling between FeI to the FeII reservoir is reduced.      

\subsection{Spectral synthesis}
\label{sect:spec}
The restricted NLTE problem, which neglects feedback effects on the atmospheric temperature and density structure, is solved using the 3D radiative transfer code \textsc{Multi3D}, developed by \citet{Botnen97} and \citet{Leenaarts09}. \citet{Amarsi16a} and Paper III describe a range of improvements recently made to the code, most importantly a new equation-of-state and background opacity package, frequency parallelisation, and improved numerical precision.  We use the same version of the code and same settings here as described in Paper III, except that we use a finer angle quadrature for the radiative transfer solution while the system converges. The Carlson A4 quadrature has 24 angles in total, four azimuthal and six inclined to the normal direction \citep{Alder63}. After the level populations have converged, the final spectrum is computed at $\mu=0.2, 0.4, 0.6, 0.8$ and 1.0, in four azimuthal directions. 

\textsc{Multi3D} calculations were performed on three atmospheric snapshots drawn from the most recent 3D radiation-hydrodynamical simulation with the \textsc{Stagger} code \citep[e.g.][]{Stein98,Collet11b,Magic13}. A detailed description of the updated simulation run will be given in a future paper (Amarsi et al. in prep.). The snapshots were resized from their original $240\times240\times230$ resolution to $60\times60\times101$, as described and tested for an earlier Solar simulation by e.g. \citet{Amarsi17}. The physical sizes of the snapshots are $6\times6\times1.5$Mm.

In addition to LTE and non-LTE line profiles computed with \textsc{Multi3D}, we computed line profiles from a larger number of 15 snapshots in LTE using \textsc{Scate} \citep{Hayek11}. Subtle differences, of the order of $2-3\%$, were noticed in the centre-to-limb behaviour of equivalent widths between the two codes, with the latter more closely resembling observations. We therefore computed our final NLTE profiles by multiplying the NLTE/LTE profile ratio found by \textsc{Multi3D} with the LTE profiles computed by \textsc{Scate}. The average effective temperature of the 15 snapshots is $5776\pm16$\,K, close enough for our purposes to the nominal $T_{\rm eff}=5772$\,K \citep{Prsa16}.

\begin{figure} 
\begin{center} 
\includegraphics[scale=0.28]{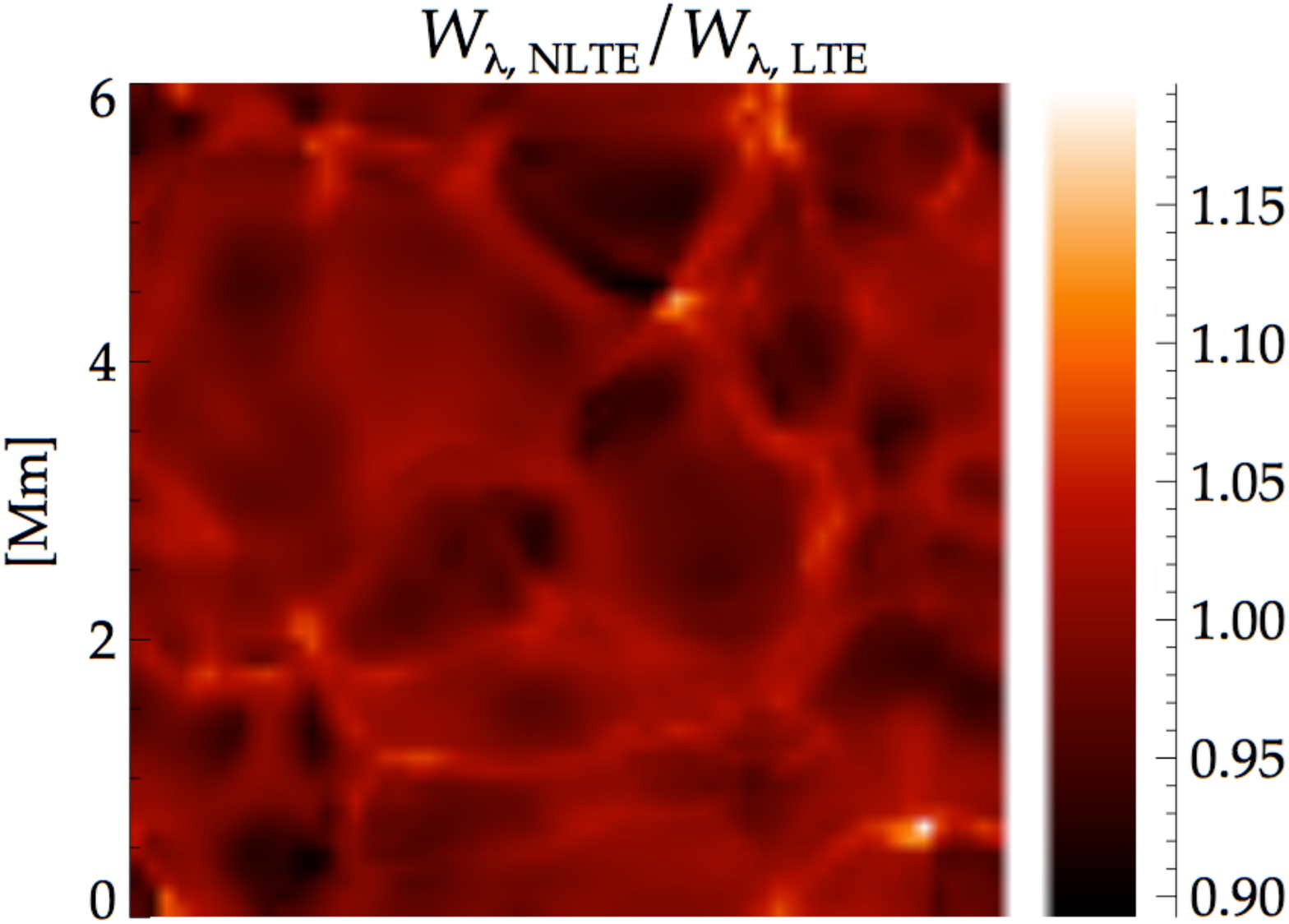}
\caption[]{The coloured image and bar on the right-hand side represent the NLTE/LTE equivalent width ratio of FeI 6151\AA\ at disk centre for a single snapshot from the solar convection simulation. In the up-flowing granules, over-ionisation causes the line to weaken in NLTE, while the inter-granular lanes display the opposite effect. The y-axis on the left-hand side indicates the spatial scale.} 
\label{fig:ratio} 
\end{center} 
\end{figure}

\section{Results and discussion}
\label{sect:results}

It is well-known that level populations of Fe do not strongly depart from LTE in the line-forming regions of the Sun and NLTE effects on line strengths are therefore small \citep[e.g.][]{Mashonkina11a,Bergemann12}.  The $\rm\langle3D\rangle$ solar model predicts significant over-ionisation of FeI to be important only at very optically thin layers ($\log(\tau_{500\rm nm})<-3.5$), while over-recombination barely dominates in deeper layers ($-2<\log(\tau_{500\rm nm})<-3$), and even deeper layers are fully thermalised. Line strengths are typically affected by less than 0.01\,dex. In full 3D, the NLTE effects vary with the convection pattern and all but the highest excited levels experience under-population in the up-flowing granules and over-population in the inter-granular lanes (Fig.\,\ref{fig:ratio}). This variation is expected given the much steeper temperature gradients of the granules and the behaviour is qualitatively similar to that found by \citet{Shchukina01}, although they predict stronger over-ionisation overall. This difference is likely due to the model atoms; our atom contains many more highly excited levels and collisions with neutral hydrogen, which strengthen the collisional coupling between FeI and the FeII reservoir and reduces NLTE effects. The surface variation can also be compared to the NLTE effects of Li\,I, Na\,I, Mg\,I and Ca\,I in metal-poor stars \citep{Asplund03,Lind13,Nordlander16}. The net effect from our 3D NLTE modelling is more over-ionisation of FeI compared to the $\rm\langle3D\rangle$ model and low-excitation lines in particular are substantially weakened. In Sect.\,\ref{sect:feabund}, we describe how these effects propagate into iron abundance corrections.    

For three FeI lines, we can compare our predicted NLTE effects with those of \citet{Holzreuter13}. Their Fig.\,8 shows histograms of the equivalent-width ratios between LTE and NLTE at each pixel in the $xy$-plane for FeI 5250\AA , 6301\AA , and 6302\AA .  For the bluer line, which has low excitation potential, we find a mean ratio of $+4$\%; significantly less than their $+15$\%. For the redder lines, we find $-1$\%, compared to their $+1$\%. Again, differences in model atom structure and adopted collisional cross-sections are most likely responsible for their stronger over-ionisation. 

\subsection{Centre-to-limb variation}
\label{sect:clv}
After performing an assessment of blends, we selected eleven iron lines, including one FeII line, within the SST wavelength ranges (See Table \ref{tab:lines}). The ten FeI lines span a wide range in wavelength and strength, but unfortunately a narrow range in lower level excitation potential. As mentioned above, low-excitation lines are most sensitive to NLTE effects, but the only observed line, 5371\AA , connected to a level below $2\,\rm eV$ in our wavelength regions is too blended to have diagnostic value and we therefore excluded it. 

The centre-to-limb behaviour is depicted in Fig.\,\ref{fig:CLVew}. The observed data points correspond to average equivalent widths measured  at each $\mu$-angle and the vertical error bars to the standard deviation of the individual pointings added to an estimated 0.5\% error due to continuum placement. Equivalent widths were measured by direct integration within wavelength ranges that were considered blend-free (see Table \ref{tab:lines}), after applying a radial velocity correction that aligns the deepest point of the line profile with the rest wavelength. The curves correspond to the predicted equivalent widths at a given abundance for each model and line, optimised to match disk-centre line strengths. The model spectra were similarly corrected to rest wavelength and integrated over the same wavelength interval as the observations. We chose this approach to enable a comparison between the models that is as fair as possible, because the 3D velocity field gives rise to a differential radial velocity effect with $\mu$ that is not captured in 1D or $\rm\langle3D\rangle$.

Full 3D modelling matches the observed centre-to-limb behaviour well; the equivalent widths are reproduced to within $\sim$5\% in both LTE and NLTE.  Comparing the two, the latter performs better for strong lines, 5367\AA\ and 5383\AA , and for the only line that becomes weaker in NLTE, 6151\AA , which has the lowest excitation potential of our lines. LTE is slightly better for 5373\AA\ and 5389\AA\ , but the differences are small and restricted to $\mu=0.2$. There is a general tendency for the 3D equivalent widths of weak lines, $W_\lambda<100\rm\,m\AA$, to be over-predicted by a few percent at $\mu=0.2$.  

We have investigated if a better match to the limb observations could be achieved by modifying the model atom. A single 3D snapshot was run with model atoms for which all hydrogen collision and electron collision rates, respectively, were reduced by an order of magnitude. The results for the atom with modified hydrogen collisions is labeled $\rm H\times0.1$ in Fig.\,\ref{fig:CLVew} and the atom with modified electron collisions is labeled $\rm e\times0.1$.  We find that reduced hydrogen collisions systematically strengthen the limb equivalent widths compared to the disk centre, such that the discrepancy with the observations increases for most lines. The effect on the level populations is such that the departures from LTE are simply shifted to deeper layers. Reducing the electron collisions also makes NLTE effects set in at deeper layers, but it also gradually enhances the over-ionisation of FeI with decreasing atmospheric depth. This has a small differential effect on the centre-to-limb variation that improves the agreement with observations in most cases. 

The change in line strength as a function of viewing angle is not well predicted by the $\rm\langle3D\rangle$ model, which gives systematically too small equivalent widths at the limb compared to the line centre. NLTE line formation alleviates the problem slightly for FeI lines, but the line strength at $\mu=0.2$ is still $5-20$\% too small. The different behaviour to full 3D modelling can be largely attributed to the treatment of velocity fields; Fig.\,\ref{fig:CLVew2} shows the results of 3D LTE modelling with the velocity field at all points and in all directions set to zero, but with a constant microturbulence of $1\rm\,km\,s^{-1}$. Evidently, this method reproduces the $\rm\langle3D\rangle$ centre-to-limb behaviour very closely, in particular for the FeII line and the high-excitation FeI lines. Fe\,I 6151\AA\ shows a slightly larger difference, which is probably caused by its lower excitation potential and thus greater sensitivity to temperature inhomogeneities. Fig.\,\ref{fig:CLVew2} also shows the results of using a 1D \textsc{MARCS} model atmosphere \citep{Gustafsson08} with $1\rm\,km\,s^{-1}.$ microturbulence, which even more strongly underestimates the the line strengths at the limb, in agreement with the Fe line analysis of \citet{Pereira09b}. The difference with respect to $\rm\langle3D\rangle$ may be attributed to the slightly steeper temperature gradient around continuum optical depth unity.

This failure of 1D models is well-known and was reported already by \citet{Holweger78}, who demonstrated that a $\mu$-dependent microturbulence may solve the problem. Their Fe line analysis found empirically that a value of $1.6\rm\,km\,s^{-1}$ is suitable at $\mu=0.3$, compared to $1.0\rm\,km\,s^{-1}$ at the disk centre, thus strengthening lines at the limb compared to centre. The same qualitative behaviour of 1D models has also been demonstrated for the centre-to-limb behaviour of the O\,I 777\,nm triplet \citep{Steffen15}. We refrain from deriving an empirical $\mu$-dependent microturbulence for 1D and $\rm\langle3D\rangle$ modelling to match our observations, but emphasize that models can now predict the 3D velocity field and thus the line broadening without invoking free parameters. We note that strengthening of lines toward the limb can occur also in 1D models as a consequence of the change in temperature gradient, without considering the velocity field, as shown e.g. for very weak ($<15\rm\,m\AA$) O\,I, Sc\,II, and Fe\,I lines by \citep{Pereira09b}.

\citet{Mashonkina13} modelled the centre-to-limb behaviour of two FeI lines, 6151\AA\ and 7780\AA , using the SST observations of \citet{Pereira09b}. They report 1D LTE modelling based on \textsc{MAFAGS-OS} model atmosphere \citep{Grupp09}, 3D LTE modelling based on a \textsc{CO$^5$BOLDT} model atmosphere \citep{Freytag12}, and $\rm\langle3D\rangle$ LTE and NLTE modelling. For the bluer line, also studied in this paper, their 1D and $\rm\langle3D\rangle$ LTE results are in good agreement with ours, but their 3D LTE modelling predicts more line strengthening toward the limb, implying that the 3D velocity field is characteristically different from our \textsc{Stagger} model. For the redder line, not studied here, they find $\rm\langle3D\rangle$ NLTE to well reproduce the centre-to-limb behaviour.

{The importance of velocity fields aside, little can be found in the literature to explain the model centre-to-limb behaviour of different lines from basic principles. In general, we find that the $\rm\langle3D\rangle$ model predicts line strengthening toward the limb for blue lines  ($<4000\AA$) and line weakening for red lines. Strong lines tend to be more weakened than weak lines at a given wavelength, as can be seen in Fig.\,\ref{fig:CLVew}. We find that this behaviour can be partly explained by equation 17.183 in \citet{Hubeny14}:

\begin{equation}
	r_\nu(\mu)\equiv I_\nu(0,\mu)/I_c(0,\mu)=[a_\nu+b_\nu\mu/(1+\beta_\nu)]/(a_\nu+b_\nu\mu)
\end{equation}

To derive this expression, the authors assume a line formed in true absorption and a linear dependence of the Planck function with continuum optical depth at a given frequency, such that $B_\nu=a_\nu+b_\nu\tau_c$, where $a_\nu$ and $b_\nu$ are positive constants. {Scattering and velocity fields are neglected. $I_\nu(0,\mu)$ is the emergent intensity at a given $\mu$-angle, $I_c(0,\mu)$ is the corresponding continuum intensity, and $\beta_\nu$ is the ratio between line and continuous opacity. Since $1+\beta_\nu>1$, the residual intensity at the limb is always higher than at disk centre for a given wavelength. Lines are therefore always predicted to weaken with decreasing $\mu$, which we have seen is true at least for red lines according to $\rm\langle3D\rangle$ modelling. It may further explain why strong lines typically decrease more in line strength than weak lines, because the inverse dependence on $\beta_\nu$ has higher influence on the residual intensity at higher $\mu$. When $\beta_\nu\gg1$, Eq. 1 approaches $a_\nu/(a_\nu+b_\nu\mu)$, which implies that the behaviour for strong lines at a given frequency is similar. This is true for the two strongest lines in our sample. Further, we estimated values for the coefficients $a_\nu$ and $b_\nu$ in the region around continuum optical depth unity for our lines and found that they change in such a way that it can explain why redder lines are more weakened than bluer (see Fig.\,\ref{fig:CLVew} and \ref{fig:CLVew2}) . However, the dependence is weaker than what the detailed modelling predicts. The validity of Eq.\,1 thus appears limited by the assumptions made.  

\begin{figure*} 
\begin{center} 
\includegraphics[scale=0.8]{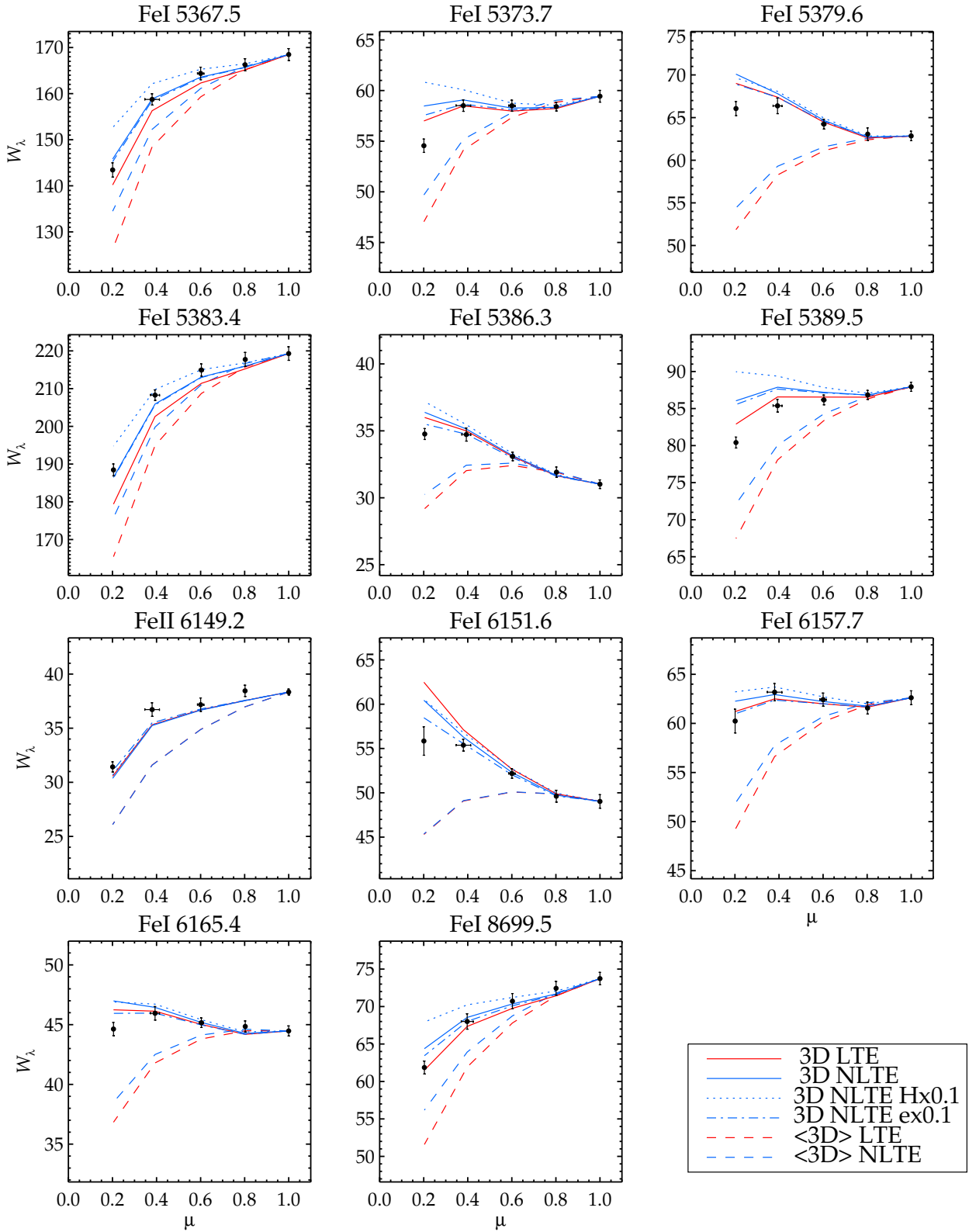}
\caption[]{Centre-to-limb variation of solar iron lines. The black bullets are observed equivalent widths and the lines represent predictions in LTE and NLTE for different model atmospheres and atomic data. A depth- and $\mu$-independent microturbulence value of $1\rm\,km\,s^{-1}$ was adopted for the $\rm\langle3D\rangle$ models. In the $\rm H\times0.1$ and $\rm e\times0.1$ models, Hydrogen and electron collisional rates were reduced by a factor ten, respectively.} 
\label{fig:CLVew} 
\end{center} 
\end{figure*}

\begin{figure*} 
\begin{center} 
\includegraphics[scale=0.8]{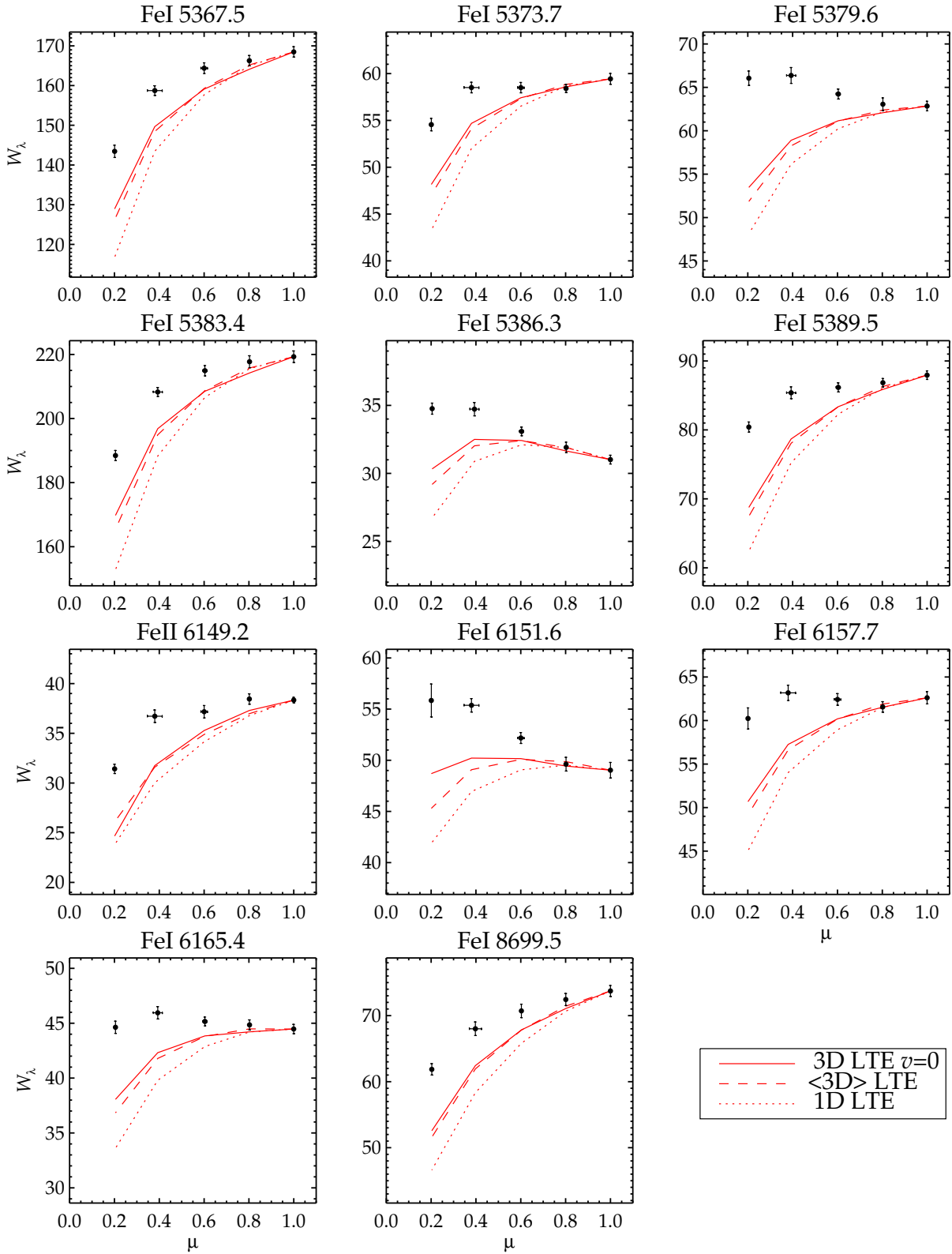}
\caption[]{The black bullets and red dashed lines are the same as in Fig.\,\ref{fig:CLVew}. The red solid lines represent 3D LTE modelling without velocity fields. For all models, we assume a depth- and $\mu$-independent microturbulence value of $1\rm\,km\,s^{-1}$. }
\label{fig:CLVew2} 
\end{center} 
\end{figure*}

\subsection{Iron abundance}
\label{sect:feabund}
\citet{Scott15} revised the solar iron abundance of \citet{Asplund09} using disk-centre intensities of 31 FeI and FeII lines, carefully selected based on blending properties, line strength, and atomic data.  They employed an earlier version of a 3D hydrodynamical \textsc{Stagger} simulation of the solar photosphere and the same Fe model atom as in Papers I and II in this series.  Abundances were first computed in 3D LTE and then corrected using NLTE calculations based on a $\rm\langle3D\rangle$ model. \citeauthor{Scott15} recommended a weighted mean abundance $\log(\epsilon_{\rm Fe})=7.47\pm0.04\rm\,dex$. They find that the excitation balance of FeI is well established in 3D LTE, whilst the NLTE abundances show a slight negative trend with excitation potential. FeI and FeII lines give a difference in mean abundance of 0.07\,dex in LTE, which decreases to 0.06\,dex after NLTE corrections have been applied. 

In this study, we re-determined iron abundances for the lines selected by \citeauthor{Scott15}, but using consistent 3D NLTE modelling. As described in the beginning of Sect.\,\ref{sect:results}, full 3D calculations predict a higher degree of over-ionisation than $\rm\langle3D\rangle$ calculations. Our new analysis technique and new model atomic data for Fe result in more positive abundance corrections for low-excitation FeI lines; between $+0.03$ and $+0.06\rm\,dex$ for $E_{\rm low}<1\rm\,eV$), while high-excitation lines ($E_{\rm low}>4\rm\,eV$) are at most affected by $-0.01$\,dex. This can be compared to $+0.11$\,dex and $+0.06$\,dex predicted for low and high-excitation lines, respectively, by \citet{Shchukina01}. Our FeI line abundances move slightly further away from fulfilling excitation balance, while the offset in ionisation balance is reduced to 0.04\,dex. The weighted mean abundance of all lines becomes slightly larger; $7.48\pm0.04$\,dex.

We repeated the model atom modifications described in Sect.\,\ref{sect:clv}, in order to evaluate if better agreement between different iron lines can be achieved. The $\rm H\times0.1$ model with altered hydrogen collisions improves neither excitation nor ionisation balance, while the $\rm e\times0.1$ model with altered electron collisions reduces the ionisation imbalance to 0.01\,dex, but at the expense of further strengthening the excitation imbalance. Turning to other potential sources of error, we note our use of electron densities computed in LTE, although important electron donors (including hydrogen) have been shown to have significant NLTE effects. We also remind the reader that the atom reduction itself may have a small impact (see Sect.\,\ref{sect:atomred}).   

Finally, \citet{Scott15} discussed the influence on FeI line abundances by magnetic fields, referencing the work of \citet{Fabbian12}, and concluded that an ionisation imbalance of order 0.02\,dex may be amended by using realistic magneto-hydrodynamic simulations with an average field strength of 100\,G \citep{TrujilloBueno04}. However, the simulations by \citet{Moore15} showed that the magnetic field must be concentrated and coherent to have an impact; a small-scale, randomly oriented field of 80\,G would not affect the iron abundance determination significantly. \citet{Shchukina15} concluded, based on the magneto-convection simulation by \citet{Rempel12}, that a small-scale dynamo with no net magnetic flux would have a typical influence on FeI line abundances of the order $+0.014$\,dex. 

\section{Conclusions}
\label{sect:conc}

We have demonstrated that full 3D, NLTE modelling of iron line formation of the Sun, using a comprehensive model atom with 463 levels, is now feasible and can successfully reproduce observed data without invoking free parameters  \citep[see also Paper III and][]{Nordlander16}. In particular we conclude here:

\begin{itemize}

\item 3D NLTE effects on low-excitation FeI lines ($<1$\,eV) are stronger than predicted by $\rm\langle3D\rangle$ modelling, resulting in 0.03-0.06\,dex higher abundances for these lines.

\item When normalised to disk-centre line strength, full 3D NLTE modelling typically over-predicts limb (here $\mu=0.2$) line strengths by approximately 5\,\%. 1D and $\rm\langle3D\rangle$ modelling in LTE and NLTE perform significantly worse, assuming a constant microturbulence of $1\rm\,km\,s^{-1}$, independent of depth and viewing angle. We stress the importance of proper treatment of the 3D velocity field for centre-to-limb modelling. 

\item The iron abundance of the Sun is found to be $\log(\epsilon_{\rm Fe})=7.48\pm0.04$\,dex, using consistent 3D NLTE modelling of the lines selected by \citet{Scott15}.

\item The ionisation imbalance between FeI and FeII line abundances in the Sun is reduced to 0.04\,dex compared to 0.06\,dex found by \citet{Scott15}. FeI line abundances show a negative slope with respect to excitation potential, similarly to metal-poor standard stars (see Paper III).

\item Rates of collisional excitation and ionisation of FeI by electrons still rely on simple semi-empirical recipes. Our tests show that less efficient electron collisions than employed in this work can improve agreement with solar observations in certain respects. This highlights the urgent need of improved data for such transitions, e.g.\ using the R-matrix method. 

\item High-quality solar observations at different viewing angles pose excellent challenges for spectral line formation models, testing the accuracy of atomic data as well as physical assumptions. Low-excitation FeI lines are of particular diagnostic importance and more data should be obtained.    

\end{itemize}

\section*{Acknowledgments} 
KL acknowledges funds from the Alexander von Humboldt Foundation in the framework of the Sofja Kovalevskaja Award endowed by the Federal Ministry of Education and Research as well as funds from the Swedish Research Council (Grant nr. 2015-00415 3) and Marie Sk\l odowska Curie Actions (Cofund Project INCA 600398). The computations were performed on resources provided by the Swedish National Infrastructure for Computing (SNIC) at UPPMAX under project p2013234. AMA and MA are supported by the Australian Research Council (grant FL110100012). PSB acknowledges support from the Royal Swedish Academy of Sciences, the Wenner-Gren Foundation, G\"oran Gustafssons Stiftelse and the Swedish Research Council. For much of this work PSB was a Royal Swedish Academy of Sciences Research Fellow supported by a grant from the Knut and Alice Wallenberg Foundation. PSB is presently partially supported by the project grant The New Milky Way from the Knut and Alice Wallenberg Foundation. Funding for the Stellar Astrophysics Centre is provided by The Danish National Research Foundation (Grant agreement no.: DNRF106). TMDP was supported by the European Research Council under the European Union's Seventh Framework Programme (FP7/2007-2013) / ERC Grant agreement No. 291058. The Swedish 1-m Solar Telescope was at the time of our observations operated on the island of La Palma by the Royal Swedish Academy of Sciences in the Spanish Observatorio del Roque de los Muchachos of the Instituto de Astrof\'{i}sica de Canarias. Finally, we thank the Max Planck Institute for Astrophysics in Garching for our SST observing time. 

\bibliographystyle{mn2emod}

\begin{thebibliography}{73}
\expandafter\ifx\csname natexlab\endcsname\relax\def\natexlab#1{#1}\fi

\bibitem[{Alder(1963)}]{Alder63}
Alder B., 1963, Methods in Computational Physics, Methods in Computational
  Physics No. v. 1. Academic Press

\bibitem[{Allen(2000)}]{Allen00}
Allen C.~W., 2000, {Allen's Astrophysical Quantities}, 4th edn. Springer,
  Berlin

\bibitem[{{Amarsi} \& {Asplund}(2017)}]{Amarsi17}
{Amarsi} A.~M., {Asplund} M., 2017, \mnras, 464, 264

\bibitem[{{Amarsi} {et~al}\mbox{.}(2016{\natexlab{a}}){Amarsi}, {Asplund},
  {Collet}, \& {Leenaarts}}]{Amarsi16a}
{Amarsi} A.~M., {Asplund} M., {Collet} R., {Leenaarts} J., 2016{\natexlab{a}},
  \mnras, 455, 3735

\bibitem[{{Amarsi} {et~al}\mbox{.}(2016{\natexlab{b}}){Amarsi}, {Lind},
  {Asplund}, {Barklem}, \& {Collet}}]{Amarsi16b}
{Amarsi} A.~M., {Lind} K., {Asplund} M., {Barklem} P.~S., {Collet} R.,
  2016{\natexlab{b}}, \mnras, 463, 1518

\bibitem[{{Anstee} \& {O'Mara}(1995)}]{Anstee95}
{Anstee} S.~D., {O'Mara} B.~J., 1995, \mnras, 276, 859

\bibitem[{{Asplund} {et~al}\mbox{.}(2003){Asplund}, {Carlsson}, \&
  {Botnen}}]{Asplund03}
{Asplund} M., {Carlsson} M., {Botnen} A.~V., 2003, \aap, 399, L31

\bibitem[{{Asplund} {et~al}\mbox{.}(2009){Asplund}, {Grevesse}, {Sauval}, \&
  {Scott}}]{Asplund09}
{Asplund} M., {Grevesse} N., {Sauval} A.~J., {Scott} P., 2009, \araa, 47, 481

\bibitem[{{Athay} \& {Lites}(1972)}]{Athay72}
{Athay} R.~G., {Lites} B.~W., 1972, \apj, 176, 809

\bibitem[{{Bailey} {et~al}\mbox{.}(2015){Bailey}, {Nagayama}, {Loisel},
  {Rochau}, {Blancard}, {Colgan}, {Cosse}, {Faussurier}, {Fontes}, {Gilleron},
  {Golovkin}, {Hansen}, {Iglesias}, {Kilcrease}, {Macfarlane}, {Mancini},
  {Nahar}, {Orban}, {Pain}, {Pradhan}, {Sherrill}, \& {Wilson}}]{Bailey15}
{Bailey} J.~E. {et~al.}, 2015, \nat, 517, 56

\bibitem[{{Bard} {et~al}\mbox{.}(1991){Bard}, {Kock}, \& {Kock}}]{BKK}
{Bard} A., {Kock} A., {Kock} M., 1991, Astron. and Astrophys., 248, 315, (BKK)

\bibitem[{{Bard} \& {Kock}(1994)}]{BK}
{Bard} A., {Kock} M., 1994, Astron. and Astrophys., 282, 1014, (BK)

\bibitem[{{Barklem}(2016)}]{Barklem16}
{Barklem} P.~S., 2016, \pra, 93, 042705

\bibitem[{{Bautista}(1997)}]{Bautista97}
{Bautista} M.~A., 1997, \aaps, 122, 167

\bibitem[{{Bautista} {et~al}\mbox{.}(2015){Bautista}, {Fivet}, {Ballance},
  {Quinet}, {Ferland}, {Mendoza}, \& {Kallman}}]{Bautista15}
{Bautista} M.~A., {Fivet} V., {Ballance} C., {Quinet} P., {Ferland} G.,
  {Mendoza} C., {Kallman} T.~R., 2015, \apj, 808, 174

\bibitem[{{Bely} \& {van Regemorter}(1970)}]{Bely70}
{Bely} O., {van Regemorter} H., 1970, \araa, 8, 329

\bibitem[{{Bergemann} {et~al}\mbox{.}(2012){Bergemann}, {Lind}, {Collet},
  {Magic}, \& {Asplund}}]{Bergemann12}
{Bergemann} M., {Lind} K., {Collet} R., {Magic} Z., {Asplund} M., 2012, \mnras,
  427, 27

\bibitem[{{Berrington} {et~al}\mbox{.}(1995){Berrington}, {Eissner}, \&
  {Norrington}}]{Berrington95}
{Berrington} K.~A., {Eissner} W.~B., {Norrington} P.~H., 1995, Computer Physics
  Communications, 92, 290

\bibitem[{{Blackwell} {et~al}\mbox{.}(1986){Blackwell}, {Booth}, {Menon}, \&
  {Petford}}]{GESB86}
{Blackwell} D.~E., {Booth} A.~J., {Menon} S.~L.~R., {Petford} A.~D., 1986,
  \mnras, 220, 289

\bibitem[{{Blackwell} {et~al}\mbox{.}(1979{\natexlab{a}}){Blackwell},
  {Ibbetson}, {Petford}, \& {Shallis}}]{BIPS}
{Blackwell} D.~E., {Ibbetson} P.~A., {Petford} A.~D., {Shallis} M.~J.,
  1979{\natexlab{a}}, \mnras, 186, 633, (BIPS)

\bibitem[{{Blackwell} {et~al}\mbox{.}(1979{\natexlab{b}}){Blackwell},
  {Petford}, \& {Shallis}}]{GESB79b}
{Blackwell} D.~E., {Petford} A.~D., {Shallis} M.~J., 1979{\natexlab{b}},
  \mnras, 186, 657

\bibitem[{{Blackwell} {et~al}\mbox{.}(1982{\natexlab{a}}){Blackwell},
  {Petford}, {Shallis}, \& {Simmons}}]{GESB82c}
{Blackwell} D.~E., {Petford} A.~D., {Shallis} M.~J., {Simmons} G.~J.,
  1982{\natexlab{a}}, \mnras, 199, 43

\bibitem[{{Blackwell} {et~al}\mbox{.}(1982{\natexlab{b}}){Blackwell},
  {Petford}, \& {Simmons}}]{GESB82d}
{Blackwell} D.~E., {Petford} A.~D., {Simmons} G.~J., 1982{\natexlab{b}},
  \mnras, 201, 595

\bibitem[{{Botnen}(1997)}]{Botnen97}
{Botnen} A., 1997, Master's thesis, Master's thesis,
  Inst.~Theor.~Astrophys.~Oslo

\bibitem[{{Collet} {et~al}\mbox{.}(2011){Collet}, {Magic}, \&
  {Asplund}}]{Collet11b}
{Collet} R., {Magic} Z., {Asplund} M., 2011, Journal of Physics Conference
  Series, 328, 012003

\bibitem[{{Den Hartog} {et~al}\mbox{.}(2014){Den Hartog}, {Ruffoni}, {Lawler},
  {Pickering}, {Lind}, \& {Brewer}}]{2014ApJS..215...23D}
{Den Hartog} E.~A., {Ruffoni} M.~P., {Lawler} J.~E., {Pickering} J.~C., {Lind}
  K., {Brewer} N.~R., 2014, \apjs, 215, 23

\bibitem[{{Fabbian} {et~al}\mbox{.}(2012){Fabbian}, {Moreno-Insertis},
  {Khomenko}, \& {Nordlund}}]{Fabbian12}
{Fabbian} D., {Moreno-Insertis} F., {Khomenko} E., {Nordlund} {\AA}., 2012,
  \aap, 548, A35

\bibitem[{{Freytag} {et~al}\mbox{.}(2012){Freytag}, {Steffen}, {Ludwig},
  {Wedemeyer-B{\"o}hm}, {Schaffenberger}, \& {Steiner}}]{Freytag12}
{Freytag} B., {Steffen} M., {Ludwig} H.-G., {Wedemeyer-B{\"o}hm} S.,
  {Schaffenberger} W., {Steiner} O., 2012, Journal of Computational Physics,
  231, 919

\bibitem[{{Fuhr} {et~al}\mbox{.}(1988){Fuhr}, {Martin}, \& {Wiese}}]{FMW}
{Fuhr} J.~R., {Martin} G.~A., {Wiese} W.~L., 1988, Journal of Physical and
  Chemical Reference Data, Volume 17, Suppl.~4.~New York: American Institute of
  Physics (AIP) and American Chemical Society, 1988, 17, (FMW)

\bibitem[{{Grupp} {et~al}\mbox{.}(2009){Grupp}, {Kurucz}, \& {Tan}}]{Grupp09}
{Grupp} F., {Kurucz} R.~L., {Tan} K., 2009, \aap, 503, 177

\bibitem[{{Gustafsson} {et~al}\mbox{.}(2008){Gustafsson}, {Edvardsson},
  {Eriksson}, {J{\o}rgensen}, {Nordlund}, \& {Plez}}]{Gustafsson08}
{Gustafsson} B., {Edvardsson} B., {Eriksson} K., {J{\o}rgensen} U.~G.,
  {Nordlund} {\AA}., {Plez} B., 2008, \aap, 486, 951

\bibitem[{{Hayek} {et~al}\mbox{.}(2011){Hayek}, {Asplund}, {Collet}, \&
  {Nordlund}}]{Hayek11}
{Hayek} W., {Asplund} M., {Collet} R., {Nordlund} {\AA}., 2011, \aap, 529, A158

\bibitem[{{Holweger} {et~al}\mbox{.}(1978){Holweger}, {Gehlsen}, \&
  {Ruland}}]{Holweger78}
{Holweger} H., {Gehlsen} M., {Ruland} F., 1978, \aap, 70, 537

\bibitem[{{Holzreuter} \& {Solanki}(2012)}]{Holzreuter12}
{Holzreuter} R., {Solanki} S.~K., 2012, \aap, 547, A46

\bibitem[{{Holzreuter} \& {Solanki}(2013)}]{Holzreuter13}
{Holzreuter} R., {Solanki} S.~K., 2013, \aap, 558, A20

\bibitem[{{Holzreuter} \& {Solanki}(2015)}]{Holzreuter15}
{Holzreuter} R., {Solanki} S.~K., 2015, \aap, 582, A101

\bibitem[{Hubeny \& Mihalas(2014)}]{Hubeny14}
Hubeny I., Mihalas D., 2014, Theory of Stellar Atmospheres: An Introduction to
  Astrophysical Non-equilibrium Quantitative Spectroscopic Analysis, Princeton
  Series in Astrophysics. Princeton University Press

\bibitem[{{Kiselman} {et~al}\mbox{.}(2011){Kiselman}, {Pereira}, {Gustafsson},
  {Asplund}, {Mel{\'e}ndez}, \& {Langhans}}]{Kiselman11}
{Kiselman} D., {Pereira} T.~M.~D., {Gustafsson} B., {Asplund} M.,
  {Mel{\'e}ndez} J., {Langhans} K., 2011, \aap, 535, A14

\bibitem[{{Korn} {et~al}\mbox{.}(2003){Korn}, {Shi}, \& {Gehren}}]{Korn03}
{Korn} A.~J., {Shi} J., {Gehren} T., 2003, \aap, 407, 691

\bibitem[{{Kurucz}(2013)}]{K13}
{Kurucz} R.~L., 2013, Robert l. kurucz on-line database of observed and
  predicted atomic transitions

\bibitem[{{Kurucz}(2014)}]{K14}
{Kurucz} R.~L., 2014, Robert l. kurucz on-line database of observed and
  predicted atomic transitions

\bibitem[{{Leenaarts} \& {Carlsson}(2009)}]{Leenaarts09}
{Leenaarts} J., {Carlsson} M., 2009, in Astronomical Society of the Pacific
  Conference Series, Vol. 415, Astronomical Society of the Pacific Conference
  Series, {B.~Lites, M.~Cheung, T.~Magara, J.~Mariska, \& K.~Reeves}, ed., pp.
  87--+

\bibitem[{{Lind} {et~al}\mbox{.}(2012){Lind}, {Bergemann}, \&
  {Asplund}}]{Lind12a}
{Lind} K., {Bergemann} M., {Asplund} M., 2012, \mnras, 427, 50

\bibitem[{{Lind} {et~al}\mbox{.}(2013){Lind}, {Melendez}, {Asplund}, {Collet},
  \& {Magic}}]{Lind13}
{Lind} K., {Melendez} J., {Asplund} M., {Collet} R., {Magic} Z., 2013, \aap,
  554, A96

\bibitem[{{Magic} {et~al}\mbox{.}(2013){Magic}, {Collet}, {Asplund},
  {Trampedach}, {Hayek}, {Chiavassa}, {Stein}, \& {Nordlund}}]{Magic13}
{Magic} Z., {Collet} R., {Asplund} M., {Trampedach} R., {Hayek} W., {Chiavassa}
  A., {Stein} R.~F., {Nordlund} {\AA}., 2013, \aap, 557, A26

\bibitem[{{Mashonkina}(2011)}]{Mashonkina11b}
{Mashonkina} L., 2011, in Magnetic Stars, pp. 314--321

\bibitem[{{Mashonkina} {et~al}\mbox{.}(2011){Mashonkina}, {Gehren}, {Shi},
  {Korn}, \& {Grupp}}]{Mashonkina11a}
{Mashonkina} L., {Gehren} T., {Shi} J.-R., {Korn} A.~J., {Grupp} F., 2011,
  \aap, 528, A87

\bibitem[{{Mashonkina} {et~al}\mbox{.}(2013){Mashonkina}, {Ludwig}, {Korn},
  {Sitnova}, \& {Caffau}}]{Mashonkina13}
{Mashonkina} L., {Ludwig} H.-G., {Korn} A., {Sitnova} T., {Caffau} E., 2013,
  Memorie della Societa Astronomica Italiana Supplementi, 24, 120

\bibitem[{{May} {et~al}\mbox{.}(1974){May}, {Richter}, \& {Wichelmann}}]{MRW}
{May} M., {Richter} J., {Wichelmann} J., 1974, \aaps, 18, 405, (MRW)

\bibitem[{{Mel{\'e}ndez} {et~al}\mbox{.}(2010){Mel{\'e}ndez}, {Casagrande},
  {Ram{\'{\i}}rez}, {Asplund}, \& {Schuster}}]{Melendez10}
{Mel{\'e}ndez} J., {Casagrande} L., {Ram{\'{\i}}rez} I., {Asplund} M.,
  {Schuster} W.~J., 2010, \aap, 515, L3+

\bibitem[{{Moore} {et~al}\mbox{.}(2015){Moore}, {Uitenbroek}, {Rempel},
  {Criscuoli}, \& {Rast}}]{Moore15}
{Moore} C.~S., {Uitenbroek} H., {Rempel} M., {Criscuoli} S., {Rast} M.~P.,
  2015, \apj, 799, 150

\bibitem[{{Nordlander} {et~al}\mbox{.}(2016){Nordlander}, {Amarsi}, {Lind},
  {Asplund}, {Barklem}, {Casey}, {Collet}, \& {Leenaarts}}]{Nordlander16}
{Nordlander} T., {Amarsi} A.~M., {Lind} K., {Asplund} M., {Barklem} P.~S.,
  {Casey} A.~R., {Collet} R., {Leenaarts} J., 2016, \aap, 597, A6

\bibitem[{{Nordlund}(1984)}]{Nordlund84}
{Nordlund} A., 1984, in Small-Scale Dynamical Processes in Quiet Stellar
  Atmospheres, {Keil} S.~L., ed., p. 181

\bibitem[{{Nordlund}(1985)}]{Nordlund85}
{Nordlund} A., 1985, in NATO Advanced Science Institutes (ASI) Series C, Vol.
  152, NATO Advanced Science Institutes (ASI) Series C, {Beckman} J.~E.,
  {Crivellari} L., eds., pp. 215--224

\bibitem[{{O'Brian} {et~al}\mbox{.}(1991){O'Brian}, {Wickliffe}, {Lawler},
  {Whaling}, \& {Brault}}]{BWL}
{O'Brian} T.~R., {Wickliffe} M.~E., {Lawler} J.~E., {Whaling} W., {Brault}
  J.~W., 1991, Journal of the Optical Society of America B Optical Physics, 8,
  1185, (BWL)

\bibitem[{{Pereira} {et~al}\mbox{.}(2009{\natexlab{a}}){Pereira}, {Asplund}, \&
  {Kiselman}}]{Pereira09b}
{Pereira} T.~M.~D., {Asplund} M., {Kiselman} D., 2009{\natexlab{a}}, \aap, 508,
  1403

\bibitem[{{Pereira} {et~al}\mbox{.}(2009{\natexlab{b}}){Pereira}, {Kiselman},
  \& {Asplund}}]{Pereira09a}
{Pereira} T.~M.~D., {Kiselman} D., {Asplund} M., 2009{\natexlab{b}}, \aap, 507,
  417

\bibitem[{{Pr{\v s}a} {et~al}\mbox{.}(2016){Pr{\v s}a}, {Harmanec}, {Torres},
  {Mamajek}, {Asplund}, {Capitaine}, {Christensen-Dalsgaard}, {Depagne},
  {Haberreiter}, {Hekker}, {Hilton}, {Kopp}, {Kostov}, {Kurtz}, {Laskar},
  {Mason}, {Milone}, {Montgomery}, {Richards}, {Schmutz}, {Schou}, \&
  {Stewart}}]{Prsa16}
{Pr{\v s}a} A. {et~al.}, 2016, \aj, 152, 41

\bibitem[{{Raassen} \& {Uylings}(1998)}]{RU}
{Raassen} A.~J.~J., {Uylings} P.~H.~M., 1998, \aap, 340, 300, (RU)

\bibitem[{{Rempel}(2014)}]{Rempel12}
{Rempel} M., 2014, \apj, 789, 132

\bibitem[{{Ruffoni} {et~al}\mbox{.}(2014){Ruffoni}, {Den Hartog}, {Lawler},
  {Brewer}, {Lind}, {Nave}, \& {Pickering}}]{2014MNRAS.441.3127R}
{Ruffoni} M.~P., {Den Hartog} E.~A., {Lawler} J.~E., {Brewer} N.~R., {Lind} K.,
  {Nave} G., {Pickering} J.~C., 2014, \mnras, 441, 3127

\bibitem[{{Rutten} \& {van der Zalm}(1984)}]{Rutten84}
{Rutten} R.~J., {van der Zalm} E.~B.~J., 1984, \aaps, 55, 143

\bibitem[{{Ryabchikova} {et~al}\mbox{.}(2015){Ryabchikova}, {Piskunov},
  {Kurucz}, {Stempels}, {Heiter}, {Pakhomov}, \& {Barklem}}]{Ryabchikova15}
{Ryabchikova} T., {Piskunov} N., {Kurucz} R.~L., {Stempels} H.~C., {Heiter} U.,
  {Pakhomov} Y., {Barklem} P.~S., 2015, \physscr, 90, 054005

\bibitem[{{Scharmer} {et~al}\mbox{.}(2003){Scharmer}, {Bjelksjo}, {Korhonen},
  {Lindberg}, \& {Petterson}}]{Scharmer03}
{Scharmer} G.~B., {Bjelksjo} K., {Korhonen} T.~K., {Lindberg} B., {Petterson}
  B., 2003, in Society of Photo-Optical Instrumentation Engineers (SPIE)
  Conference Series, Vol. 4853, Society of Photo-Optical Instrumentation
  Engineers (SPIE) Conference Series, {S.~L.~Keil \& S.~V.~Avakyan}, ed., pp.
  341--350

\bibitem[{{Scott} {et~al}\mbox{.}(2015){Scott}, {Asplund}, {Grevesse},
  {Bergemann}, \& {Sauval}}]{Scott15}
{Scott} P., {Asplund} M., {Grevesse} N., {Bergemann} M., {Sauval} A.~J., 2015,
  \aap, 573, A26

\bibitem[{{Shchukina} \& {Trujillo Bueno}(2001)}]{Shchukina01}
{Shchukina} N., {Trujillo Bueno} J., 2001, \apj, 550, 970

\bibitem[{{Shchukina} \& {Trujillo Bueno}(2015)}]{Shchukina15}
{Shchukina} N., {Trujillo Bueno} J., 2015, \aap, 579, A112

\bibitem[{{Steffen} {et~al}\mbox{.}(2015){Steffen}, {Prakapavi{\v c}ius},
  {Caffau}, {Ludwig}, {Bonifacio}, {Cayrel}, {Ku{\v c}inskas}, \&
  {Livingston}}]{Steffen15}
{Steffen} M., {Prakapavi{\v c}ius} D., {Caffau} E., {Ludwig} H.-G., {Bonifacio}
  P., {Cayrel} R., {Ku{\v c}inskas} A., {Livingston} W.~C., 2015, \aap, 583,
  A57

\bibitem[{{Stein} \& {Nordlund}(1998)}]{Stein98}
{Stein} R.~F., {Nordlund} A., 1998, \apj, 499, 914

\bibitem[{{Th{\'e}venin} \& {Idiart}(1999)}]{Thevenin99}
{Th{\'e}venin} F., {Idiart} T.~P., 1999, \apj, 521, 753

\bibitem[{{Trujillo Bueno} {et~al}\mbox{.}(2004){Trujillo Bueno}, {Shchukina},
  \& {Asensio Ramos}}]{TrujilloBueno04}
{Trujillo Bueno} J., {Shchukina} N., {Asensio Ramos} A., 2004, \nat, 430, 326

\bibitem[{{van Regemorter}(1962)}]{vanRegemorter62}
{van Regemorter} H., 1962, \apj, 136, 906

\bibitem[{{Zhang} \& {Pradhan}(1995)}]{Zhang95}
{Zhang} H.~L., {Pradhan} A.~K., 1995, \aap, 293, 953

\end{thebibliography}

\label{lastpage} \end{document}